\documentclass[aps,pra,twocolumn,showpacs,floatfix,longbibliography]{revtex4-1}

\usepackage{graphicx}
\usepackage{amssymb}
\usepackage{natbib}
\usepackage{amsmath}
\usepackage{color}
\usepackage{grffile}

\begin{document}

\title{The approach to chaos in ultracold atomic and molecular physics:\\
statistics of near-threshold bound states for Li+CaH and Li+CaF}

\author{Matthew D. Frye}
\author{Masato Morita}
\author{Christophe L. Vaillant}
\email{c.l.j.j.vaillant@durham.ac.uk}
\author{Dermot G. Green}
\altaffiliation[Present address:~]{Centre for Theoretical Atomic, Molecular and
Optical Physics, School of Mathematics and Physics, Queen's University Belfast,
Belfast BT7 1NN, Northern Ireland, United Kingdom}
\email{d.green@qub.ac.uk}
\author{Jeremy M. Hutson}
\email{Author to whom correspondence should be addressed; j.m.hutson@durham.ac.uk}
\affiliation{Joint Quantum Centre (JQC) Durham-Newcastle, Department of
Chemistry, Durham University, South Road, Durham DH1 3LE, United Kingdom.}
\date{\today}

\begin{abstract}
We calculate near-threshold bound states and Feshbach resonance positions for
atom + rigid-rotor models of the highly anisotropic systems Li+CaH and Li+CaF.
We perform statistical analysis on the resonance positions to compare with the
predictions of random matrix theory. For Li+CaH with total angular momentum
$J=0$ we find fully chaotic behavior in both the nearest-neighbor spacing
distribution and the level number variance. However, for $J>0$ we find
different behavior due to the presence of a nearly conserved quantum number.
Li+CaF ($J=0$) also shows apparently reduced levels of chaotic behavior despite
its stronger effective coupling. We suggest this may indicate the development
of another good quantum number relating to a bending motion of the complex.
However, continuously varying the rotational constant over a wide range shows
unexpected structure in the degree of chaotic behavior, including a dramatic
reduction around the rotational constant of CaF. This demonstrates the
complexity of the relationship between coupling and chaotic behavior.

\end{abstract}
\date{\today}
\maketitle

\section{Introduction}

There is currently great interest in the behavior of atoms and molecules at
ultracold temperatures. Work in this field focussed initially on simple
systems, which are amenable to detailed control by tuning with applied fields
\cite{Chin:RMP:2010}. Particular success has been achieved with ultracold
collisions and high-lying bound states of pairs of alkali-metal atoms, where a
combination of experiment and theory has produced a detailed understanding for
a variety of different systems \cite{Takekoshi:RbCs:2012, Berninger:Cs2:2013,
Julienne:Li67:2014}. This understanding has been used to study solitons in
Bose-Einstein condensates \cite{Frantzeskakis:2010}, diatomic molecule
formation \cite{Kohler:RMP:2006}, Efimov states \cite{Kraemer:2006,
Huang:2nd-Efimov:2014} and many other phenomena. However, attention is now
turning to increasingly complex systems, such as collisions of high-spin atoms
\cite{Lu:2011, Pasquiou:2012, Aikawa:2012, Baumann:2014, Maier:universal:2015,
Frisch:2015} and ground-state molecules \cite{Ni:KRb:2008,
Ospelkaus:react:2010, Parazzoli:2011, Takekoshi:RbCs:2014, Molony:RbCs:2014}.
The theory needed for a full description of these systems is often
prohibitively difficult; for example, two erbium atoms interact on 91 potential
curves \cite{Frisch:2014}, so that it is unrealistic to fit precise potentials
to experimental results in the manner that has been so successful for alkali
dimers \cite{Takekoshi:RbCs:2012, Berninger:Cs2:2013, Julienne:Li67:2014}. In
these circumstances, it is possible that the dynamics are stochastic, with
levels described by random matrix theory \cite{Mehta:rmt:1991, Guhr:1998}.
These ideas were first developed in the theory of compound nuclei
\cite{Wigner:1955, Dyson:1962}, and underlie current thinking about the nature
of quantum chaos.

A qualitative understanding of the dynamics of these highly complex systems is
important for both theorists and experimentalists. For a simple system, theory
can in principle give a complete description, but for a fully chaotic system
specific predictions are likely to be impossible. For example, it is unlikely
to be possible to map out the near-threshold bound states navigated by Feshbach
molecules except empirically \cite{Frisch:2015}. If a collision is chaotic in
nature, the collision complex will ergodically explore the entire phase-space
and the two collision partners may be trapped together for a long time. In
ultracold collisions, this may be long enough for a third body to collide with
the complex, which is likely to lead to the loss of all the particles involved
from the trap \cite{Mayle:2012}.

Random matrix theory (RMT) was first applied to ultracold collisions by
\citeauthor{Mayle:2012}, initially for atom+diatom collisions \cite{Mayle:2012}
and later for diatom+diatom collisions \cite{Mayle:2013}. The first
experimental demonstration of these ideas was provided by atom+atom collisions
for Er+Er \cite{Frisch:2014} and later Dy+Dy \cite{Maier:ChaosErDy:2015}.
Several theoretical models have also been analyzed in terms of RMT
\cite{Jachymski:ChaoticScat:2015, Frisch:2014, Maier:ChaosErDy:2015}. Within
RMT the archetypal model of quantum chaos is a matrix ensemble known as the
Gaussian Orthogonal Ensemble (GOE) \cite{Dyson:1962, Mehta:rmt:1991} and the
usual way to assess whether or not chaos is present in a level sequence is to
compare various statistics to the expectations from the GOE. In particular, the
Bohigas-Giannoni-Schmit conjecture \cite{Bohigas:1984} states that quantum
systems whose classical analogues are chaotic will show the same level
fluctuations as the GOE.

There are many statistics available to analyze level sequences (see for example
chapter 16 of \citeauthor{Mehta:rmt:1991}'s book \cite{Mehta:rmt:1991}) but
work on ultracold collisions so far has focused on two of the simplest
quantities: the distribution of nearest-neighbor spacings (NNS) and the level
number variance. In ultracold collisions, the levels (or resonances) concerned
are typically observed as a function of magnetic field rather than energy.
Erbium and dysprosium have been found to have NNS distributions and number
variances that are intermediate between those expected for chaotic and
non-chaotic systems \cite{Frisch:2014, Maier:ChaosErDy:2015}. A re-analysis of
the erbium results \cite{Mur-Petit:2015} showed that the deviations from
chaotic predictions might arise because some narrow resonances are not observed
in the experiment. A comparable degree of chaos has recently been found in
coupled-channel calculations on the remarkably simple system
Yb($^1$S)+Yb($^3$P) \cite{Green:chaos:2016}. However, similar calculations on
Li+Er found statistics consistent with a random but non-chaotic level
distribution \cite{Gonzalez-Martinez:2015}.

Atom+diatom systems provide excellent prototype systems to investigate chaotic
behavior. If vibrational excitation and electron spins are neglected, the
systems have two internal degrees of freedom; this is the minimum for classical
chaos and probably also for a quantum system to follow the predictions of RMT
\cite{Bohigas:1984}. There are well-developed formalisms for
treating collisions \cite{Bernstein:1979} and near-threshold bound states
\cite{Hutson:AMVCD:1991, Hutson:CPC:1994} in atom+diatom systems, and readily
available programs that efficiently perform the necessary calculations
\cite{molscat:v14, hutson:bound:1993}. There are many atom+diatom systems that
are of interest at low temperatures, including Rb+KRb
\cite{Ospelkaus:react:2010}, N+NH \cite{Hummon:2011} and Li+CaH
\cite{Lu:CaF:2014}.

The purpose of the present paper is to investigate quantum chaos in atom+diatom
collisions. We investigate the statistics of levels at and near threshold for
atom + rigid-rotor models of the highly anisotropic systems Li+CaH and Li+CaF,
and compare them with the predictions of RMT. For simplicity, we neglect
electron and nuclear spins. For Li+CaH with zero total angular momentum we find
excellent agreement with RMT. However for higher angular momentum we find quite
different behavior which we attribute to the presence of a nearly conserved
quantum number. Li+CaF has a larger ratio of anisotropy to rotational constant
than Li+CaH, and in this sense it is more strongly coupled, but we actually
find it is {\em less} strongly chaotic, even for zero angular momentum. This
shows that increasing coupling does not always make a system more chaotic and
may, in some circumstances, return order to the system. We analyze statistics
as a function of a continuously varying rotational constant, interpolating
between CaH and CaF and beyond, and find complex behavior with apparently
fluctuating levels of chaoticity.

\section{Potential energy surface}

The interaction between Li($^2$S) and CaH/CaF($^2\Sigma$) gives rise to
singlet and triplet electronic states of $^1$A$^\prime$ and $^3$A$^\prime$
symmetries. For sympathetic cooling in an external magnetic field, Li and
CaH/CaF would be prepared in magnetically trappable spin-stretched states, in
which all the quantum numbers for the projections of angular momentum onto the
magnetic field direction have their maximum values. Such collisions occur
primarily on the potential energy surface for the $^3$A$^\prime$ state, so we
use this surface in the present paper.

For Li+CaH, we use the {\em ab initio} $^3$A$^\prime$ interaction potential
calculated by \citeauthor{Tscherbul:2Sig:2011} \cite{Tscherbul:2Sig:2011}. For
Li+CaF, we carried out supermolecular coupled-cluster calculations of the
$^3$A$^\prime$ surface using a spin-restricted open-shell version of the
coupled-cluster method with single, double and non-iterative triple excitations
[RCCSD(T)], which is implemented in the MOLPRO package
\cite{MOLPRO:2010.1}. We used the aug-cc-pVTZ basis sets for each atom
\cite{Dunning:1989, Kendall:1992, Koput:2002}. Mid-bond functions (spdfg) were
added at the midpoint between Li and CaF. The counterpoise correction of Boys
and Bernardi \cite{Boys:1970} was used to compensate for basis set
superposition error. In all our calculations the CaF bond length was fixed at
the experimentally determined equilibrium value for the free diatom, 1.9516
{\AA} \cite{Kaledin:1999}.

\begin{figure}[t!]
\includegraphics[width=1.0\linewidth]{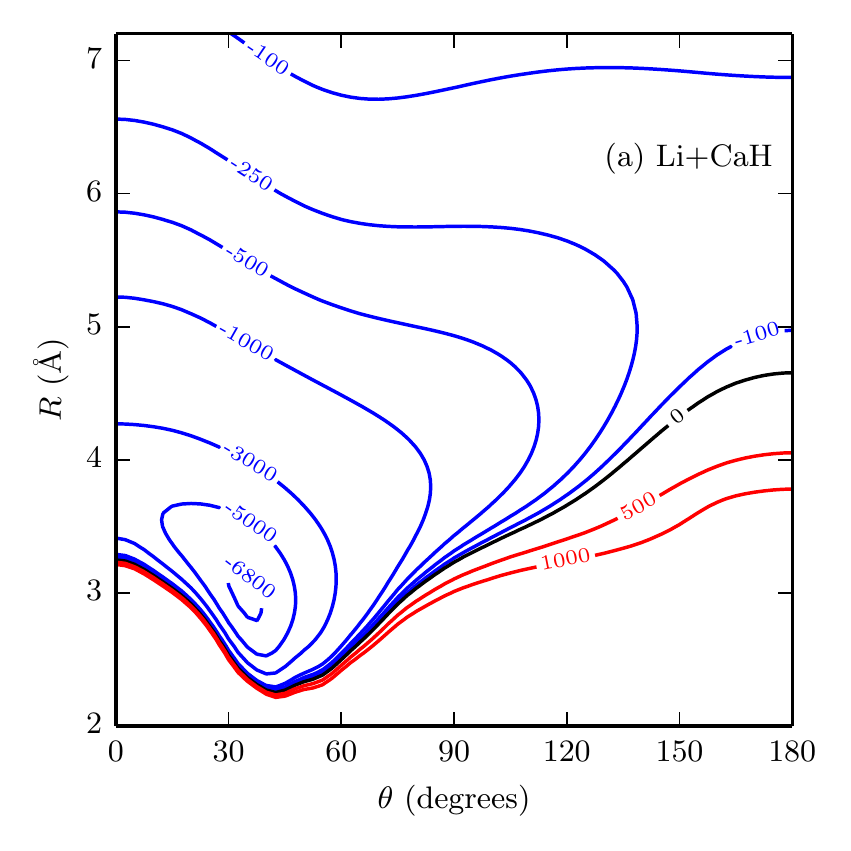}
\includegraphics[width=1.0\linewidth]{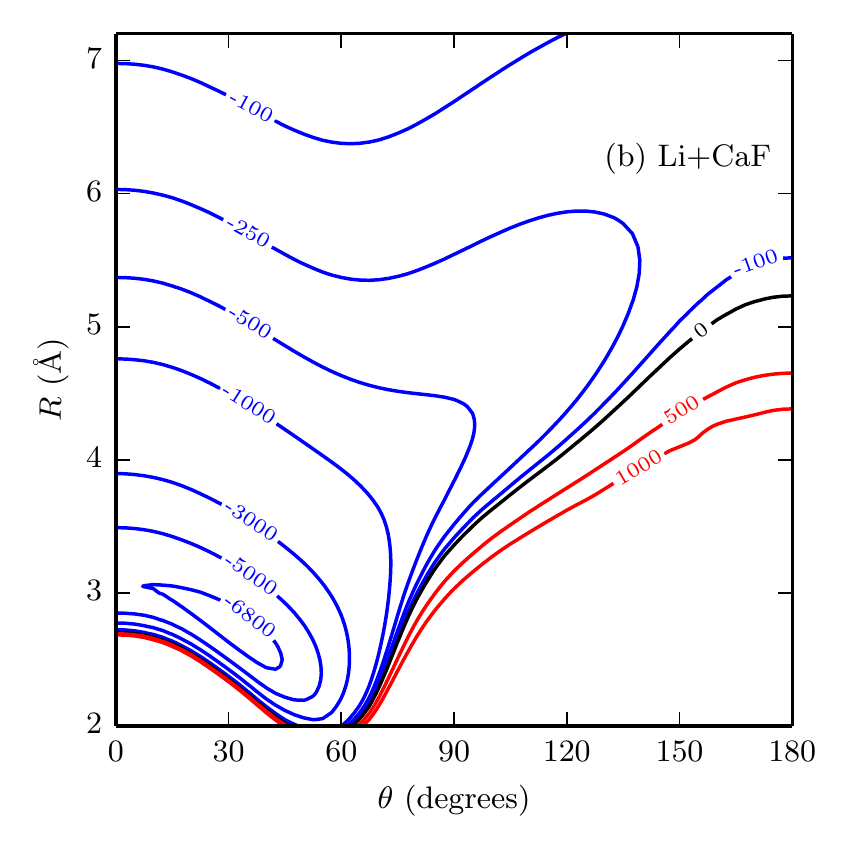}
\caption{(Color online) The ab initio $^3$A$^\prime$ interaction potential for
(a) Li+CaH and (b) Li+CaF. Contours are labeled in cm$^{-1}$. $\theta =
0^\circ$ corresponds to the Li-H(F)-Ca geometry. \label{fig:PESs}} \end{figure}

The resulting $^3$A$^\prime$ potential energy surfaces are shown in Fig.\
\ref{fig:PESs}, and it may be seen that they are strongly anisotropic, with a
deep well (7063 cm$^{-1}$ for CaH and 7258 cm$^{-1}$ for CaF) at slightly bent
Li-$X$-Ca geometries. The overall behavior of the two surfaces is similar,
though the Li+CaF interaction is slightly stronger and more anisotropic.

\section{Bound-state calculations and statistics}

\label{sec:calc_stat}

When considering whether a system is chaotic, statistical analysis would
usually be performed on a series of levels in energy with the Hamiltonian
defining the system fixed. However, the underlying Random Matrix Theory
concerns distributions taken over an ensemble of different Hamiltonians. The
assumption that the distributions over energy (for a single Hamiltonian) are
the same as those over Hamiltonians is known as spectral ergodicity
\cite{Brody:rmtreview:1981}; this is a property of RMT but not necessarily of
real systems. In recent ultracold collision studies
\cite{Jachymski:ChaoticScat:2015, Frisch:2014, Maier:ChaosErDy:2015},
statistical analysis was performed on a series of zero-energy resonance
positions as magnetic field was varied. Such experiments sample many different
Hamiltonians - albeit in a much more limited and structured way than RMT - and
so may provide a more authentic comparison with RMT than an energy spectrum
would, even if it was available. In place of the spectral ergodicity hypothesis, we now
need only to assume that different Hamiltonians are sampled in a representative
way.

Performing calculations on Li+CaH and Li+CaF in magnetic fields is a difficult
and computationally expensive task, which is beyond the scope of this paper but
will be investigated in the future. Instead, we vary the potential by an
overall scaling factor $\lambda$ \cite{Cvitas:li3:2007, Wallis:MgNH:2009,
Wallis:LiNH:2011}, which also varies the Hamiltonian but without the
theoretical complexities or computational expense associated with magnetic
fields \footnote{Note that even $\lambda=1$ should not be interpreted as the
true potential because there are significant uncertainties in the calculated
potentials.}. In this respect $\lambda$ may be considered 
a ``poor man's magnetic field".
If a system shows signs of chaos with respect to $\lambda$, it
is reasonable to label it chaotic (under the approximations made for the
dynamics).

The theory of atom-diatom interactions is well established
\cite{Bernstein:1979, Hutson:AMVCD:1991} and will not be repeated here. In the
present work, we carry out close-coupling calculations in which the diatomic
molecule is treated as a rigid rotor and its rotation $j$ is coupled to the
end-over-end rotation of the collision complex $L$ to give the total angular
momentum $J=j+L$. We use coupled-channel propagation methods
\cite{Hutson:CPC:1994} to locate bound states at threshold. We have modified
the program FIELD \cite{Hutson:field:2011} to locate bound states as a function
of $\lambda$ rather than electric or magnetic field. We use the hybrid
log-derivative propagator of \citeauthor{Alexander:1987} \cite{Alexander:1987}.
Since the diatom rotational constants, $b$, are very small ($b_{\rm
CaH}=4.2766$ cm$^{-1}$ \cite{Tscherbul:2Sig:2011} and $b_{\rm CaF}=0.339$
cm$^{-1}$ \cite{Mizushima}), we need very large basis sets of diatom rotational
functions for convergence. Unless otherwise stated, we use basis functions with
rotational quantum numbers up to $j_\text{max}=55$ and 120 for CaH and CaF,
respectively.

The real systems include diatom vibrations and electron and nuclear spins. The
harmonic frequencies for CaH and CaF are 1298 and 589 cm$^{-1}$, respectively.
Since the well depth is significantly larger than this, there will be states
from channels involving diatom vibrational excitation in the region around
threshold, although they may be sparse in energy. These are neglected in our
calculations. There will also be considerable extra density of levels due to
the spin multiplicities, although it is not clear whether the spins will be
fully involved in any chaotic dynamics or if they will be spectators. The
present results are therefore for model systems, based on the real systems but
not taking account of their full complexity.

The Gaussian Orthogonal Ensemble (GOE) is the standard RMT model for chaos in
systems with time-reversal-invariant Hamiltonians. It is a set
of $N\times N$ real symmetric matrices, with diagonal and off-diagonal elements
described by probability distributions
\begin{align}
P_{ii}(H_{ii})&=C\exp\left(-H_{ii}^2\right), \\
P_{ij}(H_{ij})&=C'\exp\left(-2H_{ij}^2\right),
\end{align}
respectively, where $C$ and $C'$ are normalization constants. The GOE has
off-diagonal elements that are on the order of the spread of the diagonal
elements and so for large $N$ are much larger than the average separation of
diagonal elements. In the context of near-threshold bound states or low-energy
collisions, this can occur when the anisotropic terms in the interaction are
comparable in magnitude to the depth of the isotropic potential, which
determines the spread of diagonal elements. The highly anisotropic potentials
of Li+CaH and Li+CaF would seem to fulfil this: the majority of the potential
well is contained in the anisotropic terms of many thousands of cm$^{-1}$,
which are significantly larger than the depth of the isotropic potential; the
latter is only 1260 cm$^{-1}$ for Li+CaH. In contrast, the Er+Er and Dy+Dy
potentials used in Ref.\ \cite{Maier:ChaosErDy:2015} have anisotropies around
10\% of their well depths \footnote{The anisotropy considered in Ref.\
\cite{Maier:ChaosErDy:2015} is based entirely on dispersion effects
\cite{Petrov:2012}. The spread of dispersion coefficients for different
potential curves of Er-Er is 10\% of their mean value, and that for Dy-Dy is
9\% of their mean value.}. In this way it is perhaps not surprising that there
are deviations from GOE predictions for Er+Er and Dy+Dy, whereas we might at
first sight expect better agreement for Li+CaH and Li+CaF.

For the statistical analysis of the levels we follow the methods of
\citeauthor{Mehta:rmt:1991} \cite{Mehta:rmt:1991} and \citeauthor{Guhr:1998}
\cite{Guhr:1998}. We denote our series of $n$ calculated level positions as
$X_i$ for $i=1,\dots,n$ \footnote{We observe only a fraction of the total
number of levels in the system, so $n \ll N$.}. First we `unfold' the level
positions to remove any systematic variation in the density and to set the
levels on a dimensionless scale with unit mean spacing. To do this we construct
the cumulative spectral function, also known as the staircase function,
\begin{equation}
S(X) = \sum_{i=1}^{n}\Theta(X-X_i),
\end{equation}
where $\Theta(x)$ is the Heaviside step function. We then fit a smooth
function, $\xi(X)$, to $S(X)$ (in this paper, fitting with a quadratic is
sufficient as the original staircase is already very nearly linear in
$\lambda$). The unfolded positions are then given as $\xi_i=\xi(X_i)$. The
nearest-neighbor spacings (NNS) are given by $s_i=\xi_{i+1}-\xi_i$ for
$i=1,2..,n-1$.

The NNS distribution is one of the main quantities of interest in statistical
analysis. In particular, for the GOE the NNS distribution is very well
approximated by the Wigner surmise \cite{Mehta:rmt:1991},
\begin{equation}
P_{\rm WD}(s)=\frac{\pi s}{2} \exp\left(-\frac{\pi s^2}{4}\right),
\end{equation}
more commonly known as the Wigner-Dyson distribution,
whereas for a random (uncorrelated) level sequence the NNS distribution
is expected to follow a Poisson distribution,
\begin{equation}
P_{\rm P}(s)=\exp\left(-s\right).
\end{equation}
Histograms of the NNS distribution provide a simple visual impression of the
statistics. The Wigner-Dyson
distribution exhibits strong level repulsion, dropping to zero at zero spacing,
whereas the Poisson distribution peaks at zero spacing. The Wigner-Dyson
distribution also falls to zero faster at large spacing than the Poisson
distribution.

Real systems do not exactly follow either $P_{\rm WD}(s)$ or $P_{\rm P}(s)$.
There are various formulas for interpolating between the two
\cite{Berry:semiclassical:1984, Izrailev:1988}. The most commonly used of
these, despite its lack of rigorous foundation, is the Brody distribution
\cite{Brody:1973},
\begin{equation}
P_{\rm B}^{(\eta)}(s)=c_\eta (\eta+1) s^\eta\exp\left(-c_\eta s^{\eta+1}\right),
\end{equation}
where
\begin{equation}
c_\eta=\left[\Gamma\left(\frac{\eta+2}{\eta+1}\right)\right]^{\eta+1},
\end{equation}
and $\eta$ is known as the Brody parameter. This is the NNS distribution that
has been used in recent publications in ultracold physics and is the one that
we use in this paper. We obtain a value of $\eta$ for a set of spacings by
maximum likelihood estimation \cite{Barlow:1989}. We maximise
the likelihood function,
\begin{equation}
L(\eta)=\prod_i P_{\rm B}^{(\eta)}(s_i),
\end{equation}
to find $\eta$. Its uncertainty is
\begin{equation}
\sigma=\left(-\frac{d^2}{d\eta^2}\ln L(\eta) \right)^{-\frac{1}{2}}.
\end{equation}
The fitted value for the Brody parameter quantifies the visual information seen
in NNS histograms: Poisson statistics yield $\eta=0$ and chaotic statistics
yield $\eta=1$. If the pattern of energy levels is in fact regular, the NNS
distribution is typically more strongly peaked than a Wigner-Dyson distribution
\cite{Berry:1977}, and so the fitted Brody parameter may be greater than unity.
We do find some fitted values $\eta>1$, but they are consistent with $\eta=1$
within statistical uncertainties.

The NNS distribution by nature captures information only about short-range
correlations but chaos is predicted to have effects over long ranges as well
\cite{Mehta:rmt:1991, Guhr:1998}. Therefore we also consider the level
number variance
\begin{equation}
\Sigma^2 (\Delta\xi)= \langle \hat{S}^2 (\Delta\xi, \xi) \rangle
- \langle \hat{S} (\Delta\xi, \xi)\rangle^2,
\end{equation}
where $\hat{S} (\Delta\xi, \xi)$ counts the number of levels in the interval
$[\xi, \xi + \Delta\xi]$ and the average is taken over the starting values
$\xi$. This characterizes the spread in the numbers of levels in intervals of
length $\Delta\xi$ and probes long-range correlations. It rises logarithmically
for the GOE, rises linearly for Poisson statistics, and oscillates around a
constant value for a regular system. While there have been some attempts to
interpolate between Poissonian and GOE behaviors of the number variance, there
is no direct analogue to the Brody distribution so we restrict ourselves to
qualitative statements about the transition between the two limiting behaviors.

\section{Results} \subsection{Li+CaH}

\begin{figure}[t!]
\includegraphics[width=0.98\linewidth]{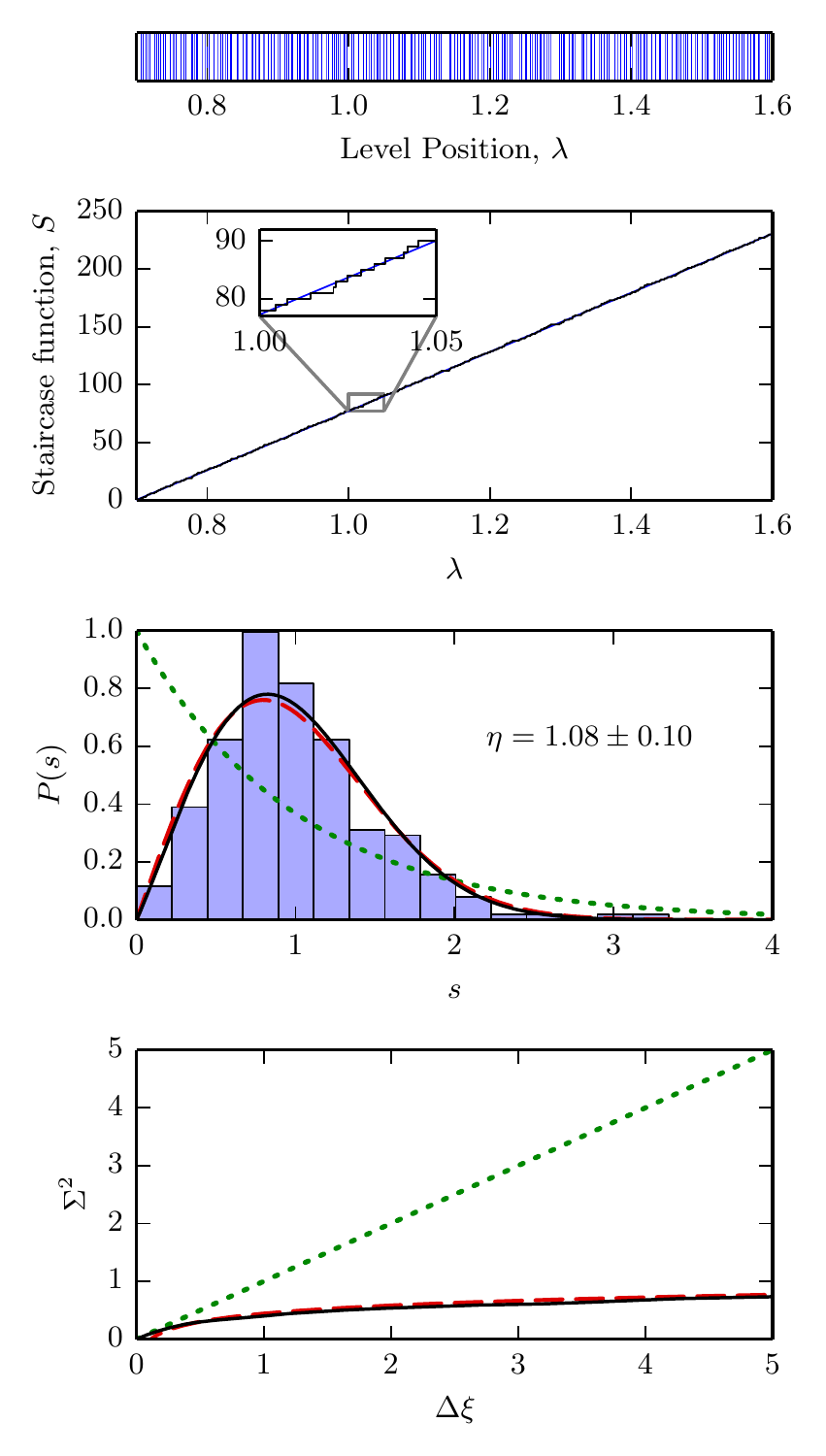}
\caption{(Color online) Statistical analysis of calculated levels of Li+CaH
($J=0$, $0.7\le \lambda \le 1.6$). From top: Level positions in $\lambda$;
Staircase function, with fitted smooth function $\xi(X)$ in blue; Histogram of
NNS distribution, green dotted and red dashed lines are Poisson and
Wigner-Dyson distributions respectively, black line is Brody distribution using
the fitted $\eta$; Level number variance, green dotted and red dashed lines are
Poisson and GOE predictions respectively. \label{fig:CaH_J0_basic}}
\end{figure}

We begin by analyzing Li+CaH for total angular momentum $J=0$ and $0.7\le
\lambda \le 1.6$. Figure \ref{fig:CaH_J0_basic} shows the calculated level
positions for $J=0$, the staircase function, a histogram of the NNS
distribution, and the level number variance. This serves as an example of the
statistical preparation and fitting described in section \ref{sec:calc_stat};
all further sequences were analyzed in the same way. The NNS distribution
clearly shows the key features of a Wigner-Dyson distribution: linear repulsion
at small spacing and a tail that dies off rapidly. The fitted Brody parameter,
$\eta=1.08\pm0.10$, is consistent with GOE predictions, and the level number
variance also follows the GOE prediction almost exactly. This Brody parameter
may be compared with values in the region 0.5 to 0.7 found for Er+Er and Dy+Dy
\cite{Maier:ChaosErDy:2015}. Our recent work on Yb+Yb$^*$
\cite{Green:chaos:2016} also found Brody parameters around 0.7 at high magnetic
field, except in cases with relatively low numbers of levels. The rigid-rotor
model of Li+CaH thus shows the clearest evidence yet found of chaotic behavior
in ultracold collisions.

\begin{figure}[t!]
\includegraphics[width=0.98\linewidth]{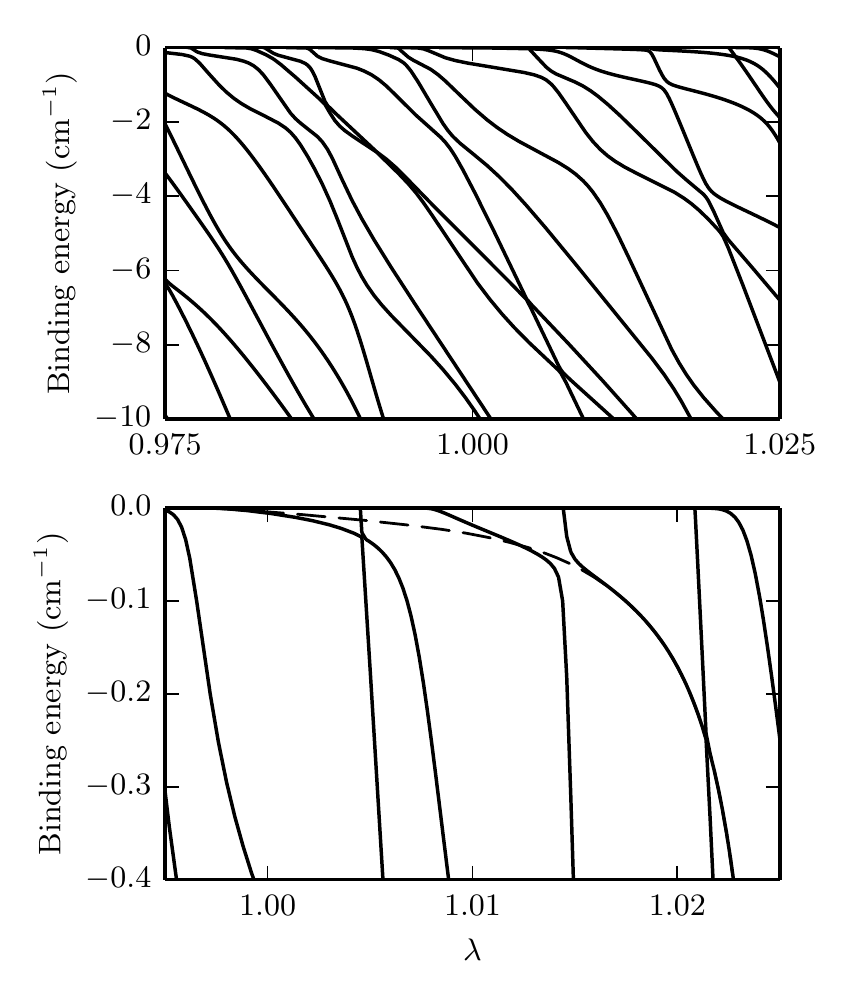}
\caption{Near-threshold bound states as a function of $\lambda$ for Li+CaH
($J=0$). The bottom panel shows an expanded view with a state approaching
threshold, crossed by several steeper states. The dashed line has been added to
follow this state through the avoided crossings and help guide the eye.
\label{fig:CaH_J0_bound}}
\end{figure}

Figure \ref{fig:CaH_J0_bound} shows near-threshold bound states for $J=0$ for a
small range of $\lambda$. The top panel shows bound states to a depth of 10
cm$^{-1}$, where the levels interact and undergo avoided crossings with a wide
variety of strengths. The lower panel is an expanded view showing a state with
a long tail curving towards threshold. This state is crossed by several steeper
states with avoided crossings of varying widths, including a crossing around
$\lambda=1.007$ that appears quite broad on this scale. Bound states with these
features have been observed experimentally in Dy+Dy
\cite{Maier:universal:2015}, although in our case the state is much more deeply
bound (by a factor of about 100 in the natural units defined by the asymptotic
van der Waals interaction relevant to such states \cite{Chin:RMP:2010}). It is
not a `halo' state because its wave function is mostly inside the outer
classical turning point \cite{Kohler:RMP:2006}, but its presence in Li+CaH
demonstrates that states with clear threshold behavior can persist across
several avoided crossings even in a system with statistics close to the
Wigner-Dyson limit.

\begin{figure}[t!]
\includegraphics[width=0.98\linewidth]{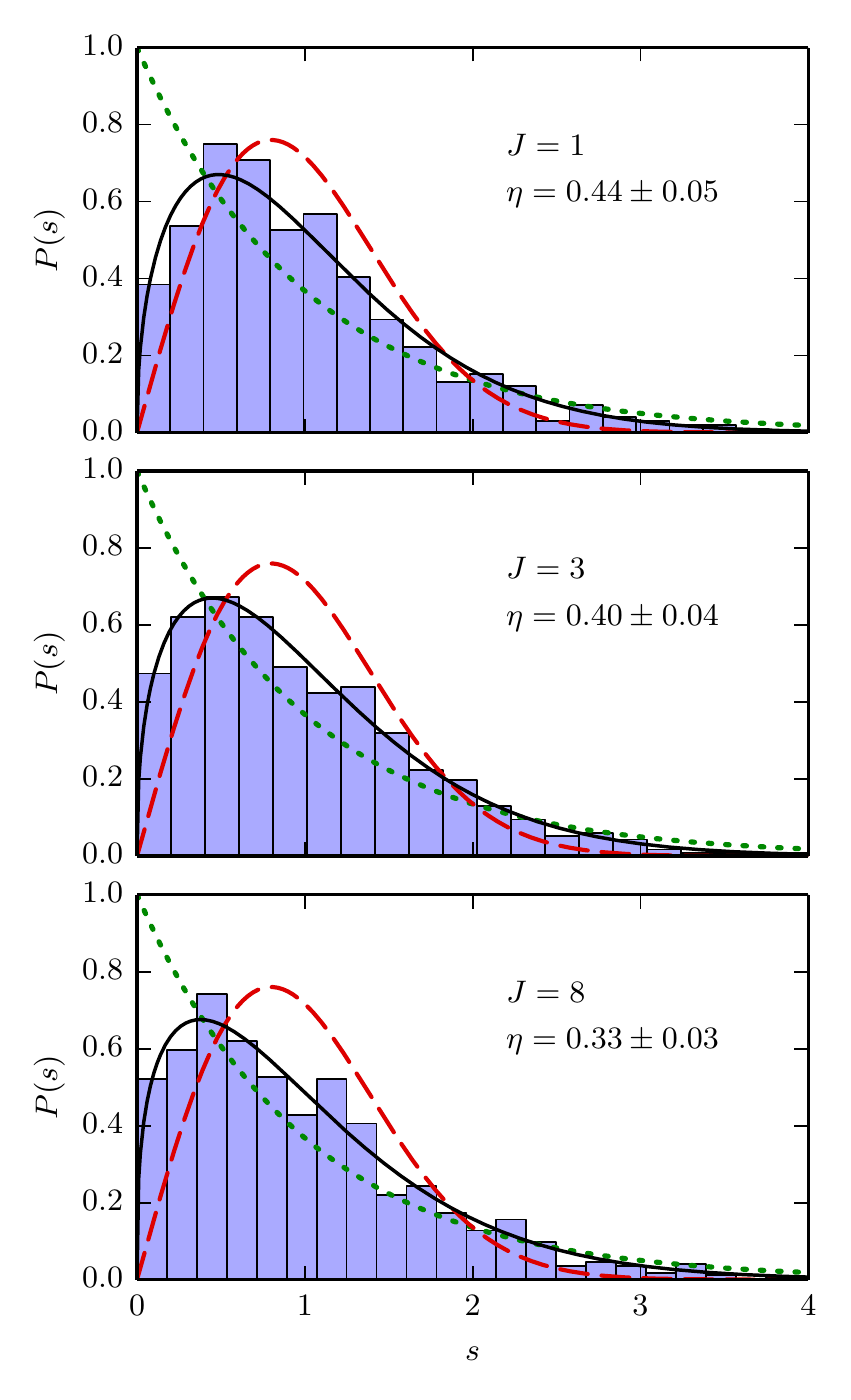}
\caption{(Color online) Nearest-neighbor spacing distributions for Li+CaH:
$J=1$ ($0.5\le \lambda \le 1.5$), $J=3$ ($0.8\le \lambda \le 1.4$), and $J=8$
($0.8\le \lambda \le 1.3$). Lines are as in Fig.\ \ref{fig:CaH_J0_basic}.
\label{fig:CaH_J-dependence}}
\end{figure}

Next we consider higher values of $J$, which correspond to higher partial waves
at the lowest rotational threshold. Figure \ref{fig:CaH_J-dependence} shows
histograms of NNS for $J$=1, 3 and 8. These notably do {\em not} show the
linear level repulsion at small spacings expected for complete chaos. Instead
there appears to be a finite probability of zero spacing. The corresponding
Brody parameters are in the region of 0.4. Although at first sight this
suggests a substantially reduced degree of chaos compared to $J=0$, such a
distribution can also occur for two overlapping but non-interacting chaotic
spectra \cite{Berry:semiclassical:1984, Rosenzweig:levelrepulsion:1960}. This
suggests that there is some form of symmetry or good quantum number present in
the system. However, we have already taken account of all rigorous symmetries,
so the quantity concerned must be only {\em approximately} conserved. On a
finer scale, the NNS distribution does indeed show some limited level
repulsion.

The nearly conserved quantum number can be understood in the body-fixed
reference frame, rather than the space-fixed frame that we use in the
coupled-channel calculations. It is the projection of the total angular
momentum $J$ (or equivalently the diatom rotation $j$) onto the intermolecular
axis, which is well known in studies of atom-diatom Van der Waals complexes
\cite{Hutson:AMVCD:1991} and is given the symbol $K$. It can take values from
$-J$ to $+J$ in integer steps. Blocks of the Hamiltonian with different $K$ are
coupled only by Coriolis terms in the body-frame representation of the
centrifugal motion; these Coriolis terms are very small compared to the
potential anisotropy in the well region, so the Hamiltonian can be considered
to be nearly block-diagonal with blocks labeled by $|K|$ and parity.

\begin{figure}[t!]
\includegraphics[width=0.98\linewidth]{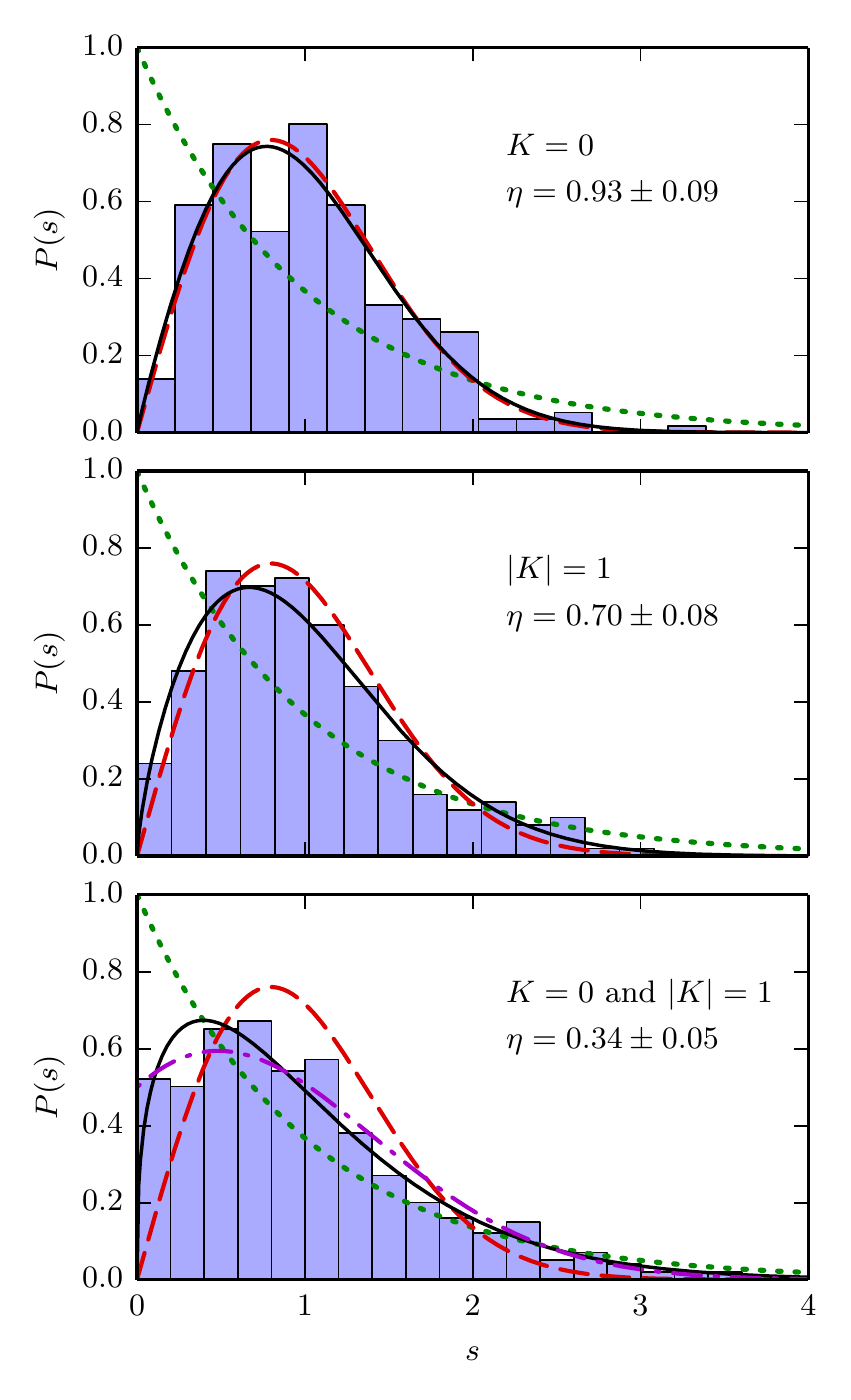}
\caption{(Color online) NNS distributions for Li+CaH ($J=1$, $0.5\le \lambda
\le 1.5$) using the helicity-decoupling approximation. $K=0$; $|K|=1$; $K=0$
and $|K|=1$. Lines are as in Fig.\ \ref{fig:CaH_J0_basic}, magenta dash-dot
line in the bottom panel is the predicted NNS distribution for two overlapping
GOEs. \label{fig:CaH_J1_helicitydecoupled}}
\end{figure}

It is possible to carry out coupled-channel calculations in the body-fixed
frame, neglecting the Coriolis terms off-diagonal in $K$. This makes the
problem block-diagonal and is known as helicity decoupling; it is often an
effective technique for calculations of atom-diatom bound states
\cite{Hutson:AMVCD:1991}. Figure \ref{fig:CaH_J1_helicitydecoupled}(a) and (b)
show NNS distributions for the separate $K=0$ and $|K|=1$ blocks. There is no
further hidden symmetry and so the NNS distributions are once again close to
the Wigner-Dyson limit. Figure \ref{fig:CaH_J1_helicitydecoupled}(c) shows the
statistics for the superposition of the two individual level sequences. This
last case is close to that of two GOE level sequences which overlap but are not
coupled. The resulting NNS distribution can be obtained from equation (3.69) of
\cite{Guhr:1998}, and is also shown in Fig.\
\ref{fig:CaH_J1_helicitydecoupled}(c). It differs from the Wigner-Dyson
distribution most obviously in that it does not vanish at zero spacing. It is
in good agreement with the results from helicity decoupling calculations and
explains the qualitative behavior of the $J\ne 0$ distributions in Fig.\ 4.

\begin{figure}[t!]
\includegraphics[width=0.98\linewidth]{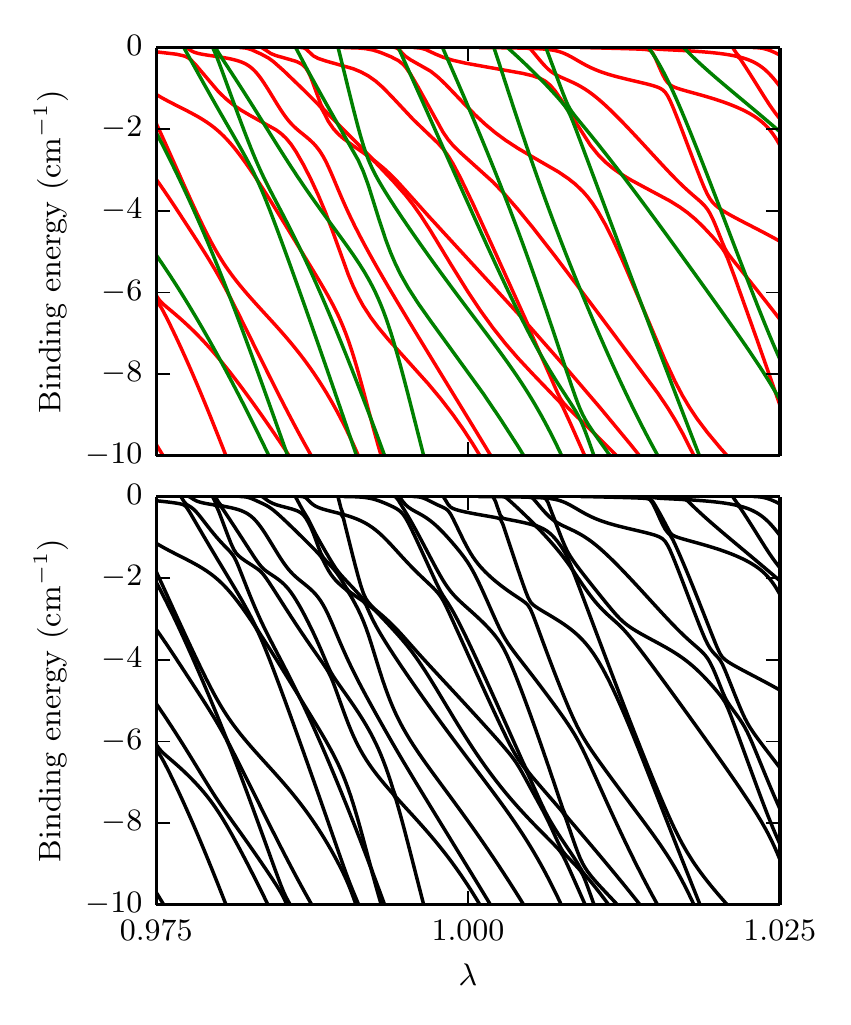}
\caption{(Color online) Near-threshold bound states as a function of $\lambda$
for Li+CaH ($J=1$): using the helicity-decoupling approximation (top; red:
$K=0$, green: $|K|=1$); full calculation (bottom).
\label{fig:CaH_J1_bound_helicitydecoupled}}
\end{figure}

The remaining differences between Fig.\ \ref{fig:CaH_J-dependence}(a) and Fig.\
\ref{fig:CaH_J1_helicitydecoupled}(c) are due to the Coriolis terms. The
quantitative effect of these terms on the statistics is beyond the scope of
this paper, but it is nevertheless informative to look at the pattern of bound
states near threshold. Figure \ref{fig:CaH_J1_bound_helicitydecoupled} shows
bound states for $J=1$ within 10 cm$^{-1}$ of threshold for a small range of
$\lambda$, both from a full calculation and within the helicity decoupling
approximation. The $K=0$ levels for $J=1$ are only slightly shifted from the
levels for $J=0$ (top panel of Fig.\ \ref{fig:CaH_J0_bound}). The $|K|=1$
levels are quite different but show the same qualitative features of many
avoided crossings of a wide variety of strengths. However, in the helicity
decoupling approximation, levels with one value of $|K|$ do not interact with
those of different $|K|$; this gives rise to a large number of true crossings,
producing an NNS distribution with finite probability at zero spacing. In the
full calculation, which takes account of the Coriolis coupling between the
blocks, the overall pattern of levels is similar but there are now narrow
avoided crossings between levels of different $|K|$. These are typically much
narrower than those between levels of the same $|K|$. This confirms our picture
of a nearly conserved quantum number, with only weak coupling between states of
different $|K|$.

\subsection{Li+CaF}
\label{sec:CaF}

The second system we consider is Li+CaF. The rotational constant for CaF is
about 13 times smaller than that for CaH, while the potential surface is quite
similar. The ratio of the anisotropy to $b$ is thus significantly greater for
CaF than for CaH. This stronger effective coupling might be expected to give
equal or greater amounts of chaos for Li+CaF compared to Li+CaH.

\begin{figure}[t!]
\includegraphics[width=0.98\linewidth]{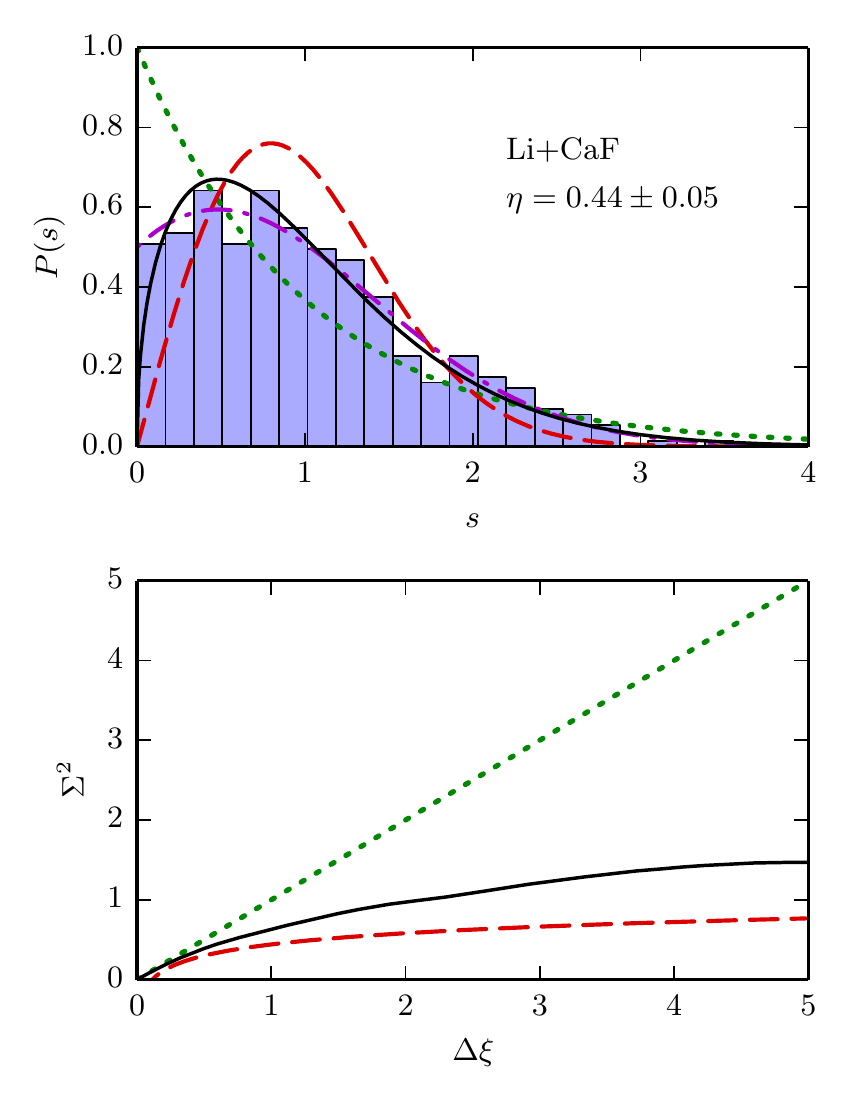}
\caption{(Color online) Calculated statistics for Li+CaF ($J=0$, $0.8\le
\lambda \le 1.4$): NNS distribution, and level number variance. Lines are as in
previous figures. \label{fig:CaF_J0}}
\end{figure}

We have performed coupled-channel calculations for Li+CaF ($J=0$) using the
same methods as for Li+CaH, but with a larger basis set because of the smaller
value of $b$. Figure \ref{fig:CaF_J0} shows the resulting level statistics.
Remarkably, the NNS distribution does not appear to show level repulsion, even
for $J=0$, but neither does it resemble a Poisson distribution; the fitted
Brody parameter is only 0.44. Once again the distribution bears a close
resemblance to the case of two overlapping but non-interacting GOE spectra as
discussed for the helicity-decoupled $J=1$ case for Li+CaH. This again hints at
the possibility of some unexpected partially good quantum number, but in this
$J=0$ case it cannot be the projection $|K|$. The level number variance for
Li+CaF ($J$=0) is also some way from the GOE prediction, although it does level
off at high spacings, in contrast to that in other near-chaotic examples
\cite{Maier:ChaosErDy:2015, Jachymski:ChaoticScat:2015}.

\begin{figure}[t!]
\includegraphics[width=1.0\linewidth]{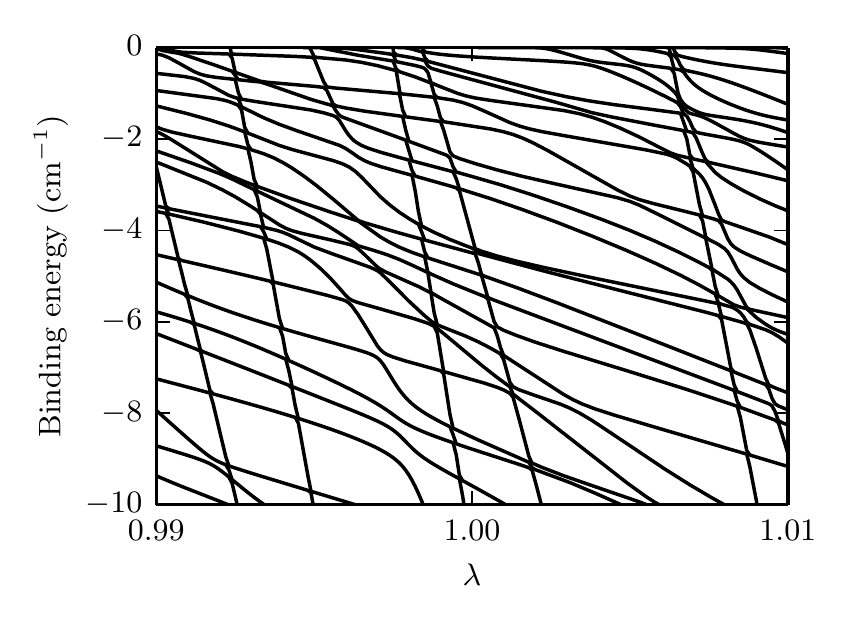}
\caption{Near-threshold bound states as a
function of $\lambda$ for Li+CaF ($J=0$). \label{fig:CaF_J0-levels}}
\end{figure}

Figure \ref{fig:CaF_J0-levels} shows the binding energies of near-threshold
levels for Li+CaF ($J=0$) as a function of $\lambda$. It may be seen that some
bound states have very steep energy gradients with respect to $\lambda$ and
that these states interact weakly with those with shallower gradients
\footnote{Although visually there may seem to be relatively few steep states
for Li+CaF, counting them with respect to $\lambda$ at fixed energy reveals
that approximately 1/3 of the levels are of this type.}. In this respect the
pattern shows a clear systematic difference from that observed for Li+CaH
(Fig.\ \ref{fig:CaH_J0_bound}), where all $J=0$ states appeared significantly
coupled and the levels could not easily be separated into classes.

There has been a great deal of work on the energy levels of atom-diatom
systems, largely aimed at understanding the dynamics of Van der Waals
complexes \cite{Hutson:AMVCD:1991}. For low anisotropies, the diatomic molecule executes hindered
rotation in the complex, and the resulting internal rotation is only weakly
coupled to the intermolecular stretching motion. However, when the effective
anisotropy is comparable to or larger than the diatom rotational constant,
there is significant mixing of diatom rotational states. As the anisotropy
increases further, the internal rotation is transformed into a bending
vibration of the triatomic molecule. Correlation diagrams showing how this
transition occurs have been given in ref.\ \onlinecite{Hutson:AMVCD:1991} for
complexes with both linear and non-linear equilibrium geometries. The low-lying
levels of a non-linear species such as Li-CaH or Li-CaF eventually execute
low-amplitude bending vibrations about their non-linear equilibrium. The degree
of mixing between bending and stretching degrees of freedom typically depends
on their relative frequencies: if the bending is either much faster or much
slower than the stretching then the modes can be separated in a
Born-Oppenheimer sense \cite{Holm77, Hutson:CBO:1980, H82RgHCl}, but if the
frequencies are comparable then they are strongly mixed.

The situation is more complicated for highly excited states, such as the
near-dissociation states that give rise to Feshbach resonances in the present
work. Some highly excited states have unstructured nodal patterns that fill the
energetically accessible space, but there are others with simple nodal patterns
that sample restricted regions of space \cite{LeSueur:1993, DePolavieja:1994,
Wright:1999, Wright:2000}. However, the paths along which such states are
localized may be complicated ones that do not correspond to obvious quantum
numbers. Because of this, it may be difficult to identify the specific nearly
conserved quantity that divides the states in Fig.\ \ref{fig:CaF_J0-levels}
into separate classes. Nevertheless, the level statistics appear to indicate
that such a quantity exists.

\subsection{Variable rotational constant}

To understand better the puzzling difference between Li+CaH and Li+CaF, we
attempt to interpolate between our two systems and extrapolate beyond them.
Since the two potential energy surfaces are so similar, we use the surface and
reduced mass for Li+CaH throughout this section, and vary the rotational
constant. We increase the number of rotational basis functions from 55 to 350
as $b$ decreases from 100 to 0.01 cm$^{-1}$ to obtain
converged level positions.

\begin{figure*}[t!]
\includegraphics[width=1.0\linewidth]{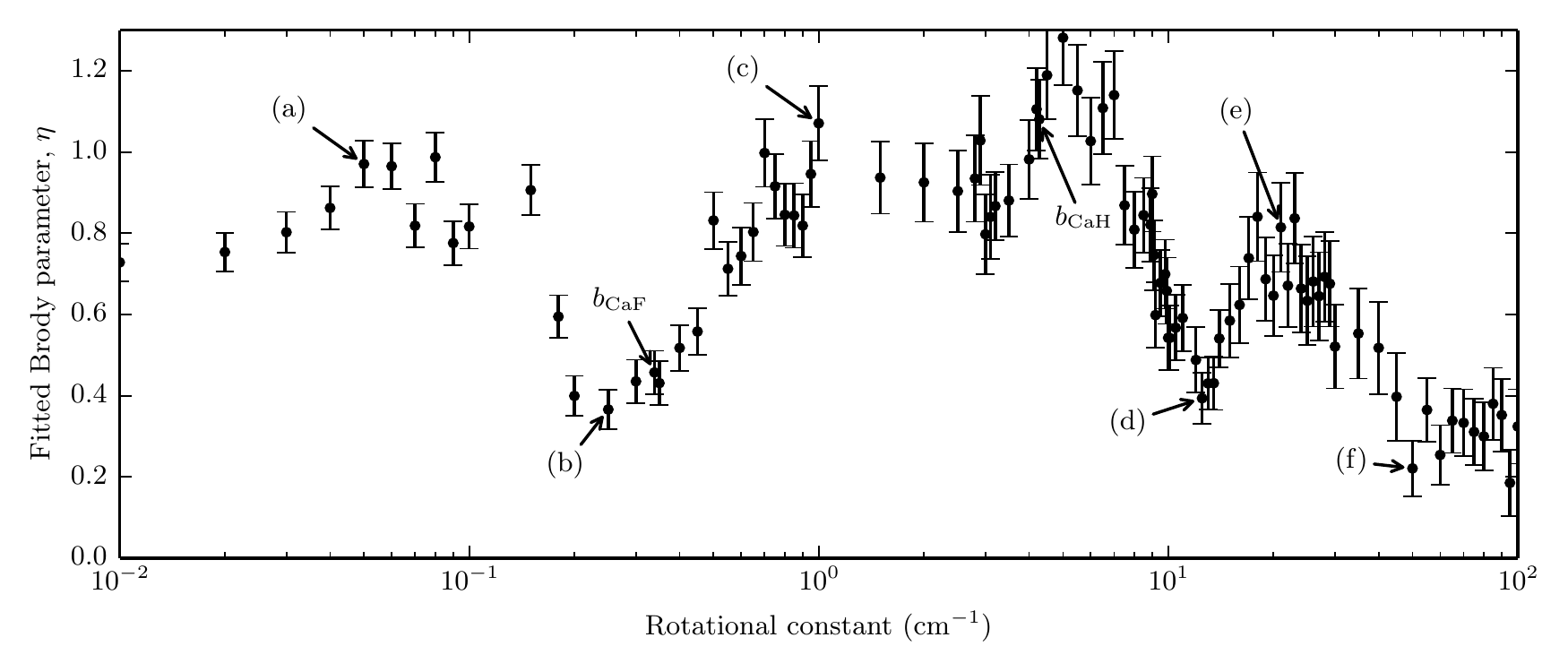}
\caption{Rotational constant dependence of fitted Brody parameter for Li+CaH
($J=0$). Labeled points are: (a) 0.05, (b) 0.25, (c) 1.0, (d) 12.5, (e) 21, (f)
50 cm$^{-1}$. $b_{\rm CaF}$ and $b_{\rm CaH}$ are also indicated.
\label{fig:brody_vs_rot}}
\end{figure*}

Figure \ref{fig:brody_vs_rot} shows the fitted Brody parameter as a function of
rotational constant. This shows an astonishing structure. It can be seen that
$b_{\rm CaH}$ lies in a relatively wide region from 0.7 to 7 cm$^{-1}$ where
$\eta$ is near unity, so that the systems can be said to be chaotic. Towards
lower rotational constant the fitted Brody parameter falls sharply to about 0.4
in the region between 0.2 and 0.4 cm$^{-1}$ -- in which $b_{\rm CaF}$ lies --
but it then rises rapidly back to values near unity for 0.05 cm$^{-1} < b <
0.08$ cm$^{-1}$ before beginning to fall slowly again. At higher values of $b$,
there is another steep and narrow trough centered around 12 cm$^{-1}$, followed
by a steady decrease towards zero as the angular and radial motions become
increasingly uncoupled.

\begin{figure*}[t!]
\includegraphics{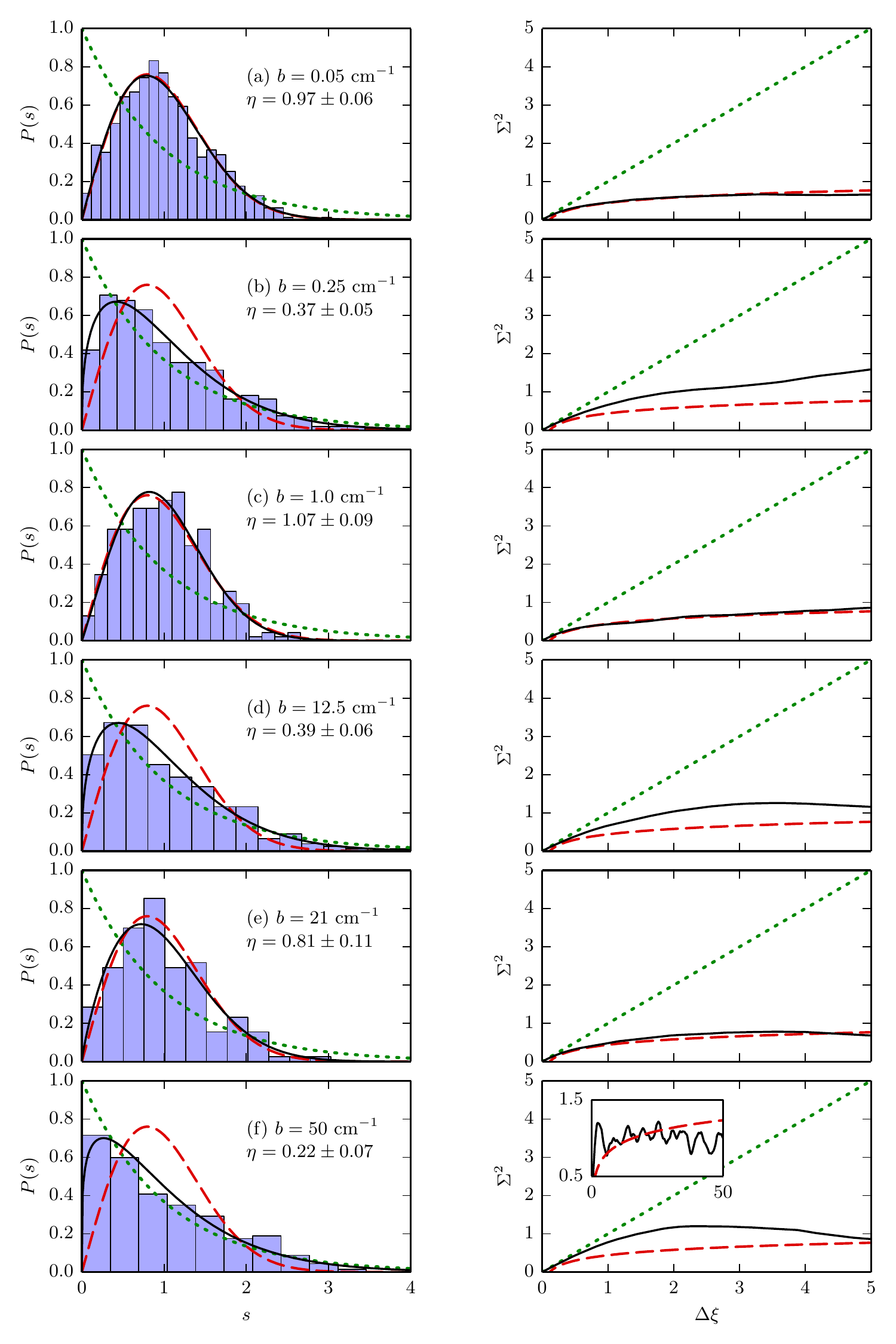}
\caption{(Color online) Examples of the NNS distribution (left) for values of
rotational constant highlighted in Fig.\ \ref{fig:brody_vs_rot}, and the
corresponding level number variance (right). Lines are as in previous figures.
\label{fig:Histo_RotDepend}}
\end{figure*}

Figure \ref{fig:Histo_RotDepend} shows statistics for sample values of the
rotational constant $b$. Those for the lowest value, $b=0.05$ cm$^{-1}$ [Fig.\
\ref{fig:Histo_RotDepend}(a)] show almost perfect agreement with the GOE
predictions for both NNS distribution and level number variance; the fitted
Brody parameter is $\eta=0.97\pm 0.06$. The second value, $b=0.25$ cm$^{-1}$,
is close to $b_{\rm CaF}=0.339$ cm$^{-1}$, but the NNS distribution [Fig.\
\ref{fig:Histo_RotDepend}(b)] differs from that seen for Li+CaF in Fig.\
\ref{fig:CaF_J0}(a), with noticeably more level repulsion despite a lower
$\eta$. $b=1.0$ cm$^{-1}$ [Fig.\ \ref{fig:Histo_RotDepend}(c)] is within the
same region of high $\eta$ as $b_{\rm CaH}$ and also shows clear signs of
chaos. $b=12.5$ cm$^{-1}$ [Fig.\ \ref{fig:Histo_RotDepend}(d)] is located in a
narrow trough of low $\eta$ and the statistics appear to be similar to case
(b). $b=21$ cm$^{-1}$ [Fig.\ \ref{fig:Histo_RotDepend}(e)] is located above the
trough in $\eta$; the Brody parameter is only 0.8 but the statistics show all
the qualitative features expected of a chaotic system. The statistics for
$b=50$ cm$^{-1}$ [Fig.\ \ref{fig:Histo_RotDepend}(f)] show an NNS distribution
that is close to Poissonian ($\eta=0.16$) because the rotational constant is
large enough for the rotational and stretching motions to be significantly
decoupled and the conditions for chaos no longer exist. However, the number
variance does not rise linearly as in the Poisson case; instead it reaches a
peak and turns downwards. The inset shows that, on a larger scale, this is the
first in a complex series of oscillations, which we attribute to the onset of
regularity and do not interpret further in this paper.

The presence of oscillations in the Brody parameter in Fig.\
\ref{fig:brody_vs_rot} is puzzling. The argument given in section \ref{sec:CaF}
above would predict a single maximum in the Brody parameter when the effective
bending and stretching frequencies are comparable for near-threshold states,
dropping off when the frequencies are very different. Fig.\
\ref{fig:brody_vs_rot} does appear to show such a maximum, but the argument
does not explain the deep minima that seem to be present either side of it. It
is possible that the flanking maxima correspond to effective bending
frequencies that match the stretching frequencies for different stretching
states, or that are rational multiples of effective stretching frequencies.

\section{Conclusions}

We have carried out calculations on threshold and near-threshold bound states
of atom + rigid-rotor models of Li+CaH and Li+CaF, and performed statistical
analysis of the resulting level sequences. For Li+CaH with zero total angular
momentum we have found the strongest signs of chaos yet observed for a
realistic ultracold collision system in either theory or experiment. However,
for non-zero total angular momentum we found a nearly good quantum number which
we identified as the body-fixed projection $K$ of the total angular momentum
$J$. The presence of this nearly conserved quantity significantly alters the
overall statistics, but the statistics for individual values of $K$ still show
chaotic behavior. The superposition of two nearly independent level sequences
in the $J=1$ case produces an NNS distribution that is distinct from the
Poisson, Wigner-Dyson and Brody distributions.

The ratio of the anisotropy to the diatom rotational constant is significantly
larger for Li+CaF than for Li+CaH. Nevertheless, contrary to expectation,
Li+CaF shows less strongly chaotic behavior even for $J=0$. The similarity of
the statistics with the case of Li+CaH ($J=1$) suggests the presence of another
nearly good quantum number in Li+CaF. The emergence of this quantum number may
be related to an adiabatic separation between a slow bending vibration and a
faster intermolecular stretch.

Finally, we have investigated how the statistics change between and beyond our
two systems by varying the rotational constant with a fixed potential. We
observe astonishing fluctuations in the levels of chaos in the system. It thus
cannot even be assumed that a system that is partway between two closely
related chaotic systems will itself be chaotic. The origin of this surprising
effect is unclear. One possibility is that stronger chaos emerges when the
bending and stretching frequencies are close to rational multiples of one
another.

This study has demonstrated that the relationship between coupling strength and
chaos is complicated. Starting from a chaotic system and increasing the
strength of a coupling does not necessarily lead to an increase in chaos. This
should not be surprising in principle: if a single term in a hamiltonian
becomes dominant, that term defines nearly good quantum numbers for the system.
The superposition of nearly independent level sequences for different values of
the nearly good quantum numbers then produces non-chaotic statistics.

It is clear that there is much to be learned from studying chaos in ultracold
collisions and high-lying bound states of atoms and molecules. Statistical
analysis can provide valuable insight when the spectra are too complex for
direct analysis. However, this study has highlighted that deviations from
chaotic behavior can be difficult to predict, even in apparently simple
systems. Future work will focus on the origins of chaotic and non-chaotic
statistics in increasingly complex systems. One question of particular
importance is whether, in real systems, all degrees of freedom are involved in
the chaotic behavior, or whether there is a hierarchy of couplings that leaves
some degrees of freedom uninvolved. Our results for Li+CaH ($J=1$) represent a
particularly simple example of a case with a clear hierarchy.

\acknowledgments

This work has been supported by the UK Engineering and Physical Sciences
Research Council (grant EP/I012044/1). The authors are grateful to Timur
Tscherbul, Jacek K\l os and Alexei Buchachenko for supplying the potential
energy surface used for Li+CaH.

\bibliography{../all,Ref_Morita}

\begin{thebibliography}{71}%
\makeatletter
\providecommand \@ifxundefined [1]{%
 \@ifx{#1\undefined}
}%
\providecommand \@ifnum [1]{%
 \ifnum #1\expandafter \@firstoftwo
 \else \expandafter \@secondoftwo
 \fi
}%
\providecommand \@ifx [1]{%
 \ifx #1\expandafter \@firstoftwo
 \else \expandafter \@secondoftwo
 \fi
}%
\providecommand \natexlab [1]{#1}%
\providecommand \enquote  [1]{``#1''}%
\providecommand \bibnamefont  [1]{#1}%
\providecommand \bibfnamefont [1]{#1}%
\providecommand \citenamefont [1]{#1}%
\providecommand \href@noop [0]{\@secondoftwo}%
\providecommand \href [0]{\begingroup \@sanitize@url \@href}%
\providecommand \@href[1]{\@@startlink{#1}\@@href}%
\providecommand \@@href[1]{\endgroup#1\@@endlink}%
\providecommand \@sanitize@url [0]{\catcode `\\12\catcode `\$12\catcode
  `\&12\catcode `\#12\catcode `\^12\catcode `\_12\catcode `\%12\relax}%
\providecommand \@@startlink[1]{}%
\providecommand \@@endlink[0]{}%
\providecommand \url  [0]{\begingroup\@sanitize@url \@url }%
\providecommand \@url [1]{\endgroup\@href {#1}{\urlprefix }}%
\providecommand \urlprefix  [0]{URL }%
\providecommand \Eprint [0]{\href }%
\providecommand \doibase [0]{http://dx.doi.org/}%
\providecommand \selectlanguage [0]{\@gobble}%
\providecommand \bibinfo  [0]{\@secondoftwo}%
\providecommand \bibfield  [0]{\@secondoftwo}%
\providecommand \translation [1]{[#1]}%
\providecommand \BibitemOpen [0]{}%
\providecommand \bibitemStop [0]{}%
\providecommand \bibitemNoStop [0]{.\EOS\space}%
\providecommand \EOS [0]{\spacefactor3000\relax}%
\providecommand \BibitemShut  [1]{\csname bibitem#1\endcsname}%
\let\auto@bib@innerbib\@empty
\bibitem [{\citenamefont {Chin}\ \emph {et~al.}(2010)\citenamefont {Chin},
  \citenamefont {Grimm}, \citenamefont {Tiesinga},\ and\ \citenamefont
  {Julienne}}]{Chin:RMP:2010}%
  \BibitemOpen
  \bibfield  {author} {\bibinfo {author} {\bibfnamefont {C.}~\bibnamefont
  {Chin}}, \bibinfo {author} {\bibfnamefont {R.}~\bibnamefont {Grimm}},
  \bibinfo {author} {\bibfnamefont {E.}~\bibnamefont {Tiesinga}}, \ and\
  \bibinfo {author} {\bibfnamefont {P.~S.}\ \bibnamefont {Julienne}},\
  }\bibfield  {title} {\enquote {\bibinfo {title} {Feshbach resonances in
  ultracold gases},}\ }\href@noop {} {\bibfield  {journal} {\bibinfo  {journal}
  {Rev. Mod. Phys.}\ }\textbf {\bibinfo {volume} {82}},\ \bibinfo {pages}
  {1225} (\bibinfo {year} {2010})}\BibitemShut {NoStop}%
\bibitem [{\citenamefont {Takekoshi}\ \emph {et~al.}(2012)\citenamefont
  {Takekoshi}, \citenamefont {Debatin}, \citenamefont {Rameshan}, \citenamefont
  {Ferlaino}, \citenamefont {Grimm}, \citenamefont {N\"agerl}, \citenamefont
  {{Le Sueur}}, \citenamefont {Hutson}, \citenamefont {Julienne}, \citenamefont
  {Kotochigova},\ and\ \citenamefont {Tiemann}}]{Takekoshi:RbCs:2012}%
  \BibitemOpen
  \bibfield  {author} {\bibinfo {author} {\bibfnamefont {T.}~\bibnamefont
  {Takekoshi}}, \bibinfo {author} {\bibfnamefont {M.}~\bibnamefont {Debatin}},
  \bibinfo {author} {\bibfnamefont {R.}~\bibnamefont {Rameshan}}, \bibinfo
  {author} {\bibfnamefont {F.}~\bibnamefont {Ferlaino}}, \bibinfo {author}
  {\bibfnamefont {R.}~\bibnamefont {Grimm}}, \bibinfo {author} {\bibfnamefont
  {H.-C.}\ \bibnamefont {N\"agerl}}, \bibinfo {author} {\bibfnamefont {C.~R.}\
  \bibnamefont {{Le Sueur}}}, \bibinfo {author} {\bibfnamefont {J.~M.}\
  \bibnamefont {Hutson}}, \bibinfo {author} {\bibfnamefont {P.~S.}\
  \bibnamefont {Julienne}}, \bibinfo {author} {\bibfnamefont {S.}~\bibnamefont
  {Kotochigova}}, \ and\ \bibinfo {author} {\bibfnamefont {E.}~\bibnamefont
  {Tiemann}},\ }\bibfield  {title} {\enquote {\bibinfo {title} {Towards the
  production of ultracold ground-state {RbCs} molecules: Feshbach resonances,
  weakly bound states, and coupled-channel models},}\ }\href@noop {} {\bibfield
   {journal} {\bibinfo  {journal} {Phys. Rev. A}\ }\textbf {\bibinfo {volume}
  {85}},\ \bibinfo {pages} {032506} (\bibinfo {year} {2012})}\BibitemShut
  {NoStop}%
\bibitem [{\citenamefont {Berninger}\ \emph {et~al.}(2013)\citenamefont
  {Berninger}, \citenamefont {Zenesini}, \citenamefont {Huang}, \citenamefont
  {Harm}, \citenamefont {N\"agerl}, \citenamefont {Ferlaino}, \citenamefont
  {Grimm}, \citenamefont {Julienne},\ and\ \citenamefont
  {Hutson}}]{Berninger:Cs2:2013}%
  \BibitemOpen
  \bibfield  {author} {\bibinfo {author} {\bibfnamefont {M.}~\bibnamefont
  {Berninger}}, \bibinfo {author} {\bibfnamefont {A.}~\bibnamefont {Zenesini}},
  \bibinfo {author} {\bibfnamefont {B.}~\bibnamefont {Huang}}, \bibinfo
  {author} {\bibfnamefont {W.}~\bibnamefont {Harm}}, \bibinfo {author}
  {\bibfnamefont {H.-C.}\ \bibnamefont {N\"agerl}}, \bibinfo {author}
  {\bibfnamefont {F.}~\bibnamefont {Ferlaino}}, \bibinfo {author}
  {\bibfnamefont {R.}~\bibnamefont {Grimm}}, \bibinfo {author} {\bibfnamefont
  {P.~S.}\ \bibnamefont {Julienne}}, \ and\ \bibinfo {author} {\bibfnamefont
  {J.~M.}\ \bibnamefont {Hutson}},\ }\bibfield  {title} {\enquote {\bibinfo
  {title} {Feshbach resonances, weakly bound molecular states and
  coupled-channel potentials for cesium at high magnetic field},}\ }\href@noop
  {} {\bibfield  {journal} {\bibinfo  {journal} {Phys. Rev. A}\ }\textbf
  {\bibinfo {volume} {87}},\ \bibinfo {pages} {032517} (\bibinfo {year}
  {2013})}\BibitemShut {NoStop}%
\bibitem [{\citenamefont {Julienne}\ and\ \citenamefont
  {Hutson}(2014)}]{Julienne:Li67:2014}%
  \BibitemOpen
  \bibfield  {author} {\bibinfo {author} {\bibfnamefont {P.~S.}\ \bibnamefont
  {Julienne}}\ and\ \bibinfo {author} {\bibfnamefont {J.~M.}\ \bibnamefont
  {Hutson}},\ }\bibfield  {title} {\enquote {\bibinfo {title} {Contrasting the
  wide {F}eshbach resonances in $^6${Li} and $^7${Li}},}\ }\href@noop {}
  {\bibfield  {journal} {\bibinfo  {journal} {Phys. Rev. A}\ }\textbf {\bibinfo
  {volume} {89}},\ \bibinfo {pages} {052715} (\bibinfo {year}
  {2014})}\BibitemShut {NoStop}%
\bibitem [{\citenamefont {Frantzeskakis}(2010)}]{Frantzeskakis:2010}%
  \BibitemOpen
  \bibfield  {author} {\bibinfo {author} {\bibfnamefont {D.~J.}\ \bibnamefont
  {Frantzeskakis}},\ }\bibfield  {title} {\enquote {\bibinfo {title} {Dark
  solitons in atomic {B}ose-{E}instein condensates: from theory to
  experiments},}\ }\href@noop {} {\bibfield  {journal} {\bibinfo  {journal} {J.
  Phys. A}\ }\textbf {\bibinfo {volume} {43}},\ \bibinfo {pages} {213001}
  (\bibinfo {year} {2010})}\BibitemShut {NoStop}%
\bibitem [{\citenamefont {K\"{o}hler}\ \emph {et~al.}(2006)\citenamefont
  {K\"{o}hler}, \citenamefont {G\'oral},\ and\ \citenamefont
  {Julienne}}]{Kohler:RMP:2006}%
  \BibitemOpen
  \bibfield  {author} {\bibinfo {author} {\bibfnamefont {T.}~\bibnamefont
  {K\"{o}hler}}, \bibinfo {author} {\bibfnamefont {K.}~\bibnamefont {G\'oral}},
  \ and\ \bibinfo {author} {\bibfnamefont {P.~S.}\ \bibnamefont {Julienne}},\
  }\bibfield  {title} {\enquote {\bibinfo {title} {Production of cold molecules
  via magnetically tunable {F}eshbach resonances},}\ }\href@noop {} {\bibfield
  {journal} {\bibinfo  {journal} {Rev. Mod. Phys.}\ }\textbf {\bibinfo {volume}
  {78}},\ \bibinfo {pages} {1311} (\bibinfo {year} {2006})}\BibitemShut
  {NoStop}%
\bibitem [{\citenamefont {Kraemer}\ \emph {et~al.}(2006)\citenamefont
  {Kraemer}, \citenamefont {Mark}, \citenamefont {Waldburger}, \citenamefont
  {Danzl}, \citenamefont {Chin}, \citenamefont {Engeser}, \citenamefont
  {Lange}, \citenamefont {Pilch}, \citenamefont {Jaakkola}, \citenamefont
  {N\"{a}gerl},\ and\ \citenamefont {Grimm}}]{Kraemer:2006}%
  \BibitemOpen
  \bibfield  {author} {\bibinfo {author} {\bibfnamefont {T.}~\bibnamefont
  {Kraemer}}, \bibinfo {author} {\bibfnamefont {M.}~\bibnamefont {Mark}},
  \bibinfo {author} {\bibfnamefont {P.}~\bibnamefont {Waldburger}}, \bibinfo
  {author} {\bibfnamefont {J.~G.}\ \bibnamefont {Danzl}}, \bibinfo {author}
  {\bibfnamefont {C.}~\bibnamefont {Chin}}, \bibinfo {author} {\bibfnamefont
  {B.}~\bibnamefont {Engeser}}, \bibinfo {author} {\bibfnamefont {A.~D.}\
  \bibnamefont {Lange}}, \bibinfo {author} {\bibfnamefont {K.}~\bibnamefont
  {Pilch}}, \bibinfo {author} {\bibfnamefont {A.}~\bibnamefont {Jaakkola}},
  \bibinfo {author} {\bibfnamefont {H.~C.}\ \bibnamefont {N\"{a}gerl}}, \ and\
  \bibinfo {author} {\bibfnamefont {R.}~\bibnamefont {Grimm}},\ }\bibfield
  {title} {\enquote {\bibinfo {title} {Evidence for {E}fimov quantum states in
  an ultracold gas of caesium atoms},}\ }\href@noop {} {\bibfield  {journal}
  {\bibinfo  {journal} {Nature}\ }\textbf {\bibinfo {volume} {440}},\ \bibinfo
  {pages} {315} (\bibinfo {year} {2006})}\BibitemShut {NoStop}%
\bibitem [{\citenamefont {Huang}\ \emph {et~al.}(2014)\citenamefont {Huang},
  \citenamefont {Sidorenkov}, \citenamefont {Grimm},\ and\ \citenamefont
  {Hutson}}]{Huang:2nd-Efimov:2014}%
  \BibitemOpen
  \bibfield  {author} {\bibinfo {author} {\bibfnamefont {B.}~\bibnamefont
  {Huang}}, \bibinfo {author} {\bibfnamefont {L.~A.}\ \bibnamefont
  {Sidorenkov}}, \bibinfo {author} {\bibfnamefont {R.}~\bibnamefont {Grimm}}, \
  and\ \bibinfo {author} {\bibfnamefont {J.~M.}\ \bibnamefont {Hutson}},\
  }\bibfield  {title} {\enquote {\bibinfo {title} {Observation of the second
  triatomic resonance in {Efimov}'s scenario},}\ }\href@noop {} {\bibfield
  {journal} {\bibinfo  {journal} {Phys. Rev. Lett.}\ }\textbf {\bibinfo
  {volume} {112}},\ \bibinfo {pages} {190401} (\bibinfo {year}
  {2014})}\BibitemShut {NoStop}%
\bibitem [{\citenamefont {Lu}\ \emph {et~al.}(2011)\citenamefont {Lu},
  \citenamefont {Burdick}, \citenamefont {Youn},\ and\ \citenamefont
  {Lev}}]{Lu:2011}%
  \BibitemOpen
  \bibfield  {author} {\bibinfo {author} {\bibfnamefont {M.}~\bibnamefont
  {Lu}}, \bibinfo {author} {\bibfnamefont {N.~Q.}\ \bibnamefont {Burdick}},
  \bibinfo {author} {\bibfnamefont {S.~H.}\ \bibnamefont {Youn}}, \ and\
  \bibinfo {author} {\bibfnamefont {B.~L.}\ \bibnamefont {Lev}},\ }\bibfield
  {title} {\enquote {\bibinfo {title} {Strongly dipolar {B}ose-{E}instein
  condensate of dysprosium},}\ }\href {\doibase 10.1103/PhysRevLett.107.190401}
  {\bibfield  {journal} {\bibinfo  {journal} {Phys. Rev. Lett.}\ }\textbf
  {\bibinfo {volume} {107}},\ \bibinfo {pages} {190401} (\bibinfo {year}
  {2011})}\BibitemShut {NoStop}%
\bibitem [{\citenamefont {Pasquiou}\ \emph {et~al.}(2012)\citenamefont
  {Pasquiou}, \citenamefont {Mar\'echal}, \citenamefont {Vernac}, \citenamefont
  {Gorceix},\ and\ \citenamefont {Laburthe-Tolra}}]{Pasquiou:2012}%
  \BibitemOpen
  \bibfield  {author} {\bibinfo {author} {\bibfnamefont {B.}~\bibnamefont
  {Pasquiou}}, \bibinfo {author} {\bibfnamefont {E.}~\bibnamefont
  {Mar\'echal}}, \bibinfo {author} {\bibfnamefont {L.}~\bibnamefont {Vernac}},
  \bibinfo {author} {\bibfnamefont {O.}~\bibnamefont {Gorceix}}, \ and\
  \bibinfo {author} {\bibfnamefont {B.}~\bibnamefont {Laburthe-Tolra}},\
  }\bibfield  {title} {\enquote {\bibinfo {title} {Thermodynamics of a
  {B}ose-{E}instein condensate with free magnetization},}\ }\href {\doibase
  10.1103/PhysRevLett.108.045307} {\bibfield  {journal} {\bibinfo  {journal}
  {Phys. Rev. Lett.}\ }\textbf {\bibinfo {volume} {108}},\ \bibinfo {pages}
  {045307} (\bibinfo {year} {2012})}\BibitemShut {NoStop}%
\bibitem [{\citenamefont {Aikawa}\ \emph {et~al.}(2012)\citenamefont {Aikawa},
  \citenamefont {Frisch}, \citenamefont {Mark}, \citenamefont {Baier},
  \citenamefont {Rietzler}, \citenamefont {Grimm},\ and\ \citenamefont
  {Ferlaino}}]{Aikawa:2012}%
  \BibitemOpen
  \bibfield  {author} {\bibinfo {author} {\bibfnamefont {K.}~\bibnamefont
  {Aikawa}}, \bibinfo {author} {\bibfnamefont {A.}~\bibnamefont {Frisch}},
  \bibinfo {author} {\bibfnamefont {M.}~\bibnamefont {Mark}}, \bibinfo {author}
  {\bibfnamefont {S.}~\bibnamefont {Baier}}, \bibinfo {author} {\bibfnamefont
  {A.}~\bibnamefont {Rietzler}}, \bibinfo {author} {\bibfnamefont
  {R.}~\bibnamefont {Grimm}}, \ and\ \bibinfo {author} {\bibfnamefont
  {F.}~\bibnamefont {Ferlaino}},\ }\bibfield  {title} {\enquote {\bibinfo
  {title} {Bose-{E}instein condensation of erbium},}\ }\href {\doibase
  10.1103/PhysRevLett.108.210401} {\bibfield  {journal} {\bibinfo  {journal}
  {Phys. Rev. Lett.}\ }\textbf {\bibinfo {volume} {108}},\ \bibinfo {pages}
  {210401} (\bibinfo {year} {2012})}\BibitemShut {NoStop}%
\bibitem [{\citenamefont {Baumann}\ \emph {et~al.}(2014)\citenamefont
  {Baumann}, \citenamefont {Burdick}, \citenamefont {Lu},\ and\ \citenamefont
  {Lev}}]{Baumann:2014}%
  \BibitemOpen
  \bibfield  {author} {\bibinfo {author} {\bibfnamefont {K.}~\bibnamefont
  {Baumann}}, \bibinfo {author} {\bibfnamefont {N.~Q.}\ \bibnamefont
  {Burdick}}, \bibinfo {author} {\bibfnamefont {M.}~\bibnamefont {Lu}}, \ and\
  \bibinfo {author} {\bibfnamefont {B.~L.}\ \bibnamefont {Lev}},\ }\bibfield
  {title} {\enquote {\bibinfo {title} {Observation of low-field
  {F}ano-{F}eshbach resonances in ultracold gases of dysprosium},}\ }\href
  {\doibase 10.1103/PhysRevA.89.020701} {\bibfield  {journal} {\bibinfo
  {journal} {Phys. Rev. A}\ }\textbf {\bibinfo {volume} {89}},\ \bibinfo
  {pages} {020701} (\bibinfo {year} {2014})}\BibitemShut {NoStop}%
\bibitem [{\citenamefont {Maier}\ \emph
  {et~al.}(2015{\natexlab{a}})\citenamefont {Maier}, \citenamefont
  {Ferrier-Barbut}, \citenamefont {Kadau}, \citenamefont {Schmitt},
  \citenamefont {Wenzel}, \citenamefont {Wink}, \citenamefont {Pfau},
  \citenamefont {Jachymski},\ and\ \citenamefont
  {Julienne}}]{Maier:universal:2015}%
  \BibitemOpen
  \bibfield  {author} {\bibinfo {author} {\bibfnamefont {T.}~\bibnamefont
  {Maier}}, \bibinfo {author} {\bibfnamefont {I.}~\bibnamefont
  {Ferrier-Barbut}}, \bibinfo {author} {\bibfnamefont {H.}~\bibnamefont
  {Kadau}}, \bibinfo {author} {\bibfnamefont {M.}~\bibnamefont {Schmitt}},
  \bibinfo {author} {\bibfnamefont {M.}~\bibnamefont {Wenzel}}, \bibinfo
  {author} {\bibfnamefont {C.}~\bibnamefont {Wink}}, \bibinfo {author}
  {\bibfnamefont {T.}~\bibnamefont {Pfau}}, \bibinfo {author} {\bibfnamefont
  {K.}~\bibnamefont {Jachymski}}, \ and\ \bibinfo {author} {\bibfnamefont
  {P.~S.}\ \bibnamefont {Julienne}},\ }\bibfield  {title} {\enquote {\bibinfo
  {title} {Broad universal {F}eshbach resonances in the chaotic spectrum of
  dysprosium atoms},}\ }\href {\doibase 10.1103/PhysRevA.92.060702} {\bibfield
  {journal} {\bibinfo  {journal} {Phys. Rev. A}\ }\textbf {\bibinfo {volume}
  {92}},\ \bibinfo {pages} {060702} (\bibinfo {year}
  {2015}{\natexlab{a}})}\BibitemShut {NoStop}%
\bibitem [{\citenamefont {Frisch}\ \emph {et~al.}(2015)\citenamefont {Frisch},
  \citenamefont {Mark}, \citenamefont {Aikawa}, \citenamefont {Baier},
  \citenamefont {Grimm}, \citenamefont {Petrov}, \citenamefont {Kotochigova},
  \citenamefont {Qu\'em\'ener}, \citenamefont {Lepers}, \citenamefont
  {Dulieu},\ and\ \citenamefont {Ferlaino}}]{Frisch:2015}%
  \BibitemOpen
  \bibfield  {author} {\bibinfo {author} {\bibfnamefont {A.}~\bibnamefont
  {Frisch}}, \bibinfo {author} {\bibfnamefont {M.}~\bibnamefont {Mark}},
  \bibinfo {author} {\bibfnamefont {K.}~\bibnamefont {Aikawa}}, \bibinfo
  {author} {\bibfnamefont {S.}~\bibnamefont {Baier}}, \bibinfo {author}
  {\bibfnamefont {R.}~\bibnamefont {Grimm}}, \bibinfo {author} {\bibfnamefont
  {A.}~\bibnamefont {Petrov}}, \bibinfo {author} {\bibfnamefont
  {S.}~\bibnamefont {Kotochigova}}, \bibinfo {author} {\bibfnamefont
  {G.}~\bibnamefont {Qu\'em\'ener}}, \bibinfo {author} {\bibfnamefont
  {M.}~\bibnamefont {Lepers}}, \bibinfo {author} {\bibfnamefont
  {O.}~\bibnamefont {Dulieu}}, \ and\ \bibinfo {author} {\bibfnamefont
  {F.}~\bibnamefont {Ferlaino}},\ }\bibfield  {title} {\enquote {\bibinfo
  {title} {Ultracold dipolar molecules composed of strongly magnetic atoms},}\
  }\href {\doibase 10.1103/PhysRevLett.115.203201} {\bibfield  {journal}
  {\bibinfo  {journal} {Phys. Rev. Lett.}\ }\textbf {\bibinfo {volume} {115}},\
  \bibinfo {pages} {203201} (\bibinfo {year} {2015})}\BibitemShut {NoStop}%
\bibitem [{\citenamefont {Ni}\ \emph {et~al.}(2008)\citenamefont {Ni},
  \citenamefont {Ospelkaus}, \citenamefont {{de Miranda}}, \citenamefont
  {Pe'er}, \citenamefont {Neyenhuis}, \citenamefont {Zirbel}, \citenamefont
  {Kotochigova}, \citenamefont {Julienne}, \citenamefont {Jin},\ and\
  \citenamefont {Ye}}]{Ni:KRb:2008}%
  \BibitemOpen
  \bibfield  {author} {\bibinfo {author} {\bibfnamefont {K.-K.}\ \bibnamefont
  {Ni}}, \bibinfo {author} {\bibfnamefont {S.}~\bibnamefont {Ospelkaus}},
  \bibinfo {author} {\bibfnamefont {M.~H.~G.}\ \bibnamefont {{de Miranda}}},
  \bibinfo {author} {\bibfnamefont {A.}~\bibnamefont {Pe'er}}, \bibinfo
  {author} {\bibfnamefont {B.}~\bibnamefont {Neyenhuis}}, \bibinfo {author}
  {\bibfnamefont {J.~J.}\ \bibnamefont {Zirbel}}, \bibinfo {author}
  {\bibfnamefont {S.}~\bibnamefont {Kotochigova}}, \bibinfo {author}
  {\bibfnamefont {P.~S.}\ \bibnamefont {Julienne}}, \bibinfo {author}
  {\bibfnamefont {D.~S.}\ \bibnamefont {Jin}}, \ and\ \bibinfo {author}
  {\bibfnamefont {J.}~\bibnamefont {Ye}},\ }\bibfield  {title} {\enquote
  {\bibinfo {title} {A high phase-space-density gas of polar molecules in the
  rovibrational ground state},}\ }\href@noop {} {\bibfield  {journal} {\bibinfo
   {journal} {Science}\ }\textbf {\bibinfo {volume} {322}},\ \bibinfo {pages}
  {231} (\bibinfo {year} {2008})}\BibitemShut {NoStop}%
\bibitem [{\citenamefont {Ospelkaus}\ \emph {et~al.}(2010)\citenamefont
  {Ospelkaus}, \citenamefont {Ni}, \citenamefont {Wang}, \citenamefont {{de
  Miranda}}, \citenamefont {Neyenhuis}, \citenamefont {Qu\'{e}m\'{e}ner},
  \citenamefont {Julienne}, \citenamefont {Bohn}, \citenamefont {Jin},\ and\
  \citenamefont {Ye}}]{Ospelkaus:react:2010}%
  \BibitemOpen
  \bibfield  {author} {\bibinfo {author} {\bibfnamefont {S.}~\bibnamefont
  {Ospelkaus}}, \bibinfo {author} {\bibfnamefont {K.-K.}\ \bibnamefont {Ni}},
  \bibinfo {author} {\bibfnamefont {D.}~\bibnamefont {Wang}}, \bibinfo {author}
  {\bibfnamefont {M.~H.~G.}\ \bibnamefont {{de Miranda}}}, \bibinfo {author}
  {\bibfnamefont {B.}~\bibnamefont {Neyenhuis}}, \bibinfo {author}
  {\bibfnamefont {G.}~\bibnamefont {Qu\'{e}m\'{e}ner}}, \bibinfo {author}
  {\bibfnamefont {P.~S.}\ \bibnamefont {Julienne}}, \bibinfo {author}
  {\bibfnamefont {J.~L.}\ \bibnamefont {Bohn}}, \bibinfo {author}
  {\bibfnamefont {D.~S.}\ \bibnamefont {Jin}}, \ and\ \bibinfo {author}
  {\bibfnamefont {J.}~\bibnamefont {Ye}},\ }\bibfield  {title} {\enquote
  {\bibinfo {title} {Quantum-state controlled chemical reactions of ultracold
  {KRb} molecules},}\ }\href@noop {} {\bibfield  {journal} {\bibinfo  {journal}
  {Science}\ }\textbf {\bibinfo {volume} {327}},\ \bibinfo {pages} {853}
  (\bibinfo {year} {2010})}\BibitemShut {NoStop}%
\bibitem [{\citenamefont {Parazzoli}\ \emph {et~al.}(2011)\citenamefont
  {Parazzoli}, \citenamefont {Fitch}, \citenamefont {\.Zuchowski},
  \citenamefont {Hutson},\ and\ \citenamefont {Lewandowski}}]{Parazzoli:2011}%
  \BibitemOpen
  \bibfield  {author} {\bibinfo {author} {\bibfnamefont {L.~P.}\ \bibnamefont
  {Parazzoli}}, \bibinfo {author} {\bibfnamefont {N.~J.}\ \bibnamefont
  {Fitch}}, \bibinfo {author} {\bibfnamefont {P.~S.}\ \bibnamefont
  {\.Zuchowski}}, \bibinfo {author} {\bibfnamefont {J.~M.}\ \bibnamefont
  {Hutson}}, \ and\ \bibinfo {author} {\bibfnamefont {H.~J.}\ \bibnamefont
  {Lewandowski}},\ }\bibfield  {title} {\enquote {\bibinfo {title} {Large
  effects of electric fields on atom-molecule collisions at millikelvin
  temperatures},}\ }\href@noop {} {\bibfield  {journal} {\bibinfo  {journal}
  {Phys. Rev. Lett.}\ }\textbf {\bibinfo {volume} {106}},\ \bibinfo {pages}
  {193201} (\bibinfo {year} {2011})}\BibitemShut {NoStop}%
\bibitem [{\citenamefont {Takekoshi}\ \emph {et~al.}(2014)\citenamefont
  {Takekoshi}, \citenamefont {Reichs\"ollner}, \citenamefont {Schindewolf},
  \citenamefont {Hutson}, \citenamefont {{Le Sueur}}, \citenamefont {Dulieu},
  \citenamefont {Ferlaino}, \citenamefont {Grimm},\ and\ \citenamefont
  {N\"agerl}}]{Takekoshi:RbCs:2014}%
  \BibitemOpen
  \bibfield  {author} {\bibinfo {author} {\bibfnamefont {T.}~\bibnamefont
  {Takekoshi}}, \bibinfo {author} {\bibfnamefont {L.}~\bibnamefont
  {Reichs\"ollner}}, \bibinfo {author} {\bibfnamefont {A.}~\bibnamefont
  {Schindewolf}}, \bibinfo {author} {\bibfnamefont {J.~M.}\ \bibnamefont
  {Hutson}}, \bibinfo {author} {\bibfnamefont {C.~R.}\ \bibnamefont {{Le
  Sueur}}}, \bibinfo {author} {\bibfnamefont {O.}~\bibnamefont {Dulieu}},
  \bibinfo {author} {\bibfnamefont {F.}~\bibnamefont {Ferlaino}}, \bibinfo
  {author} {\bibfnamefont {R.}~\bibnamefont {Grimm}}, \ and\ \bibinfo {author}
  {\bibfnamefont {H.-C.}\ \bibnamefont {N\"agerl}},\ }\bibfield  {title}
  {\enquote {\bibinfo {title} {Ultracold dense samples of dipolar {RbCs}
  molecules in the rovibrational and hyperfine ground state},}\ }\href@noop {}
  {\bibfield  {journal} {\bibinfo  {journal} {Phys. Rev. Lett.}\ }\textbf
  {\bibinfo {volume} {113}},\ \bibinfo {pages} {205301} (\bibinfo {year}
  {2014})}\BibitemShut {NoStop}%
\bibitem [{\citenamefont {Molony}\ \emph {et~al.}(2014)\citenamefont {Molony},
  \citenamefont {Gregory}, \citenamefont {Ji}, \citenamefont {Lu},
  \citenamefont {K\"oppinger}, \citenamefont {{Le Sueur}}, \citenamefont
  {Blackley}, \citenamefont {Hutson},\ and\ \citenamefont
  {Cornish}}]{Molony:RbCs:2014}%
  \BibitemOpen
  \bibfield  {author} {\bibinfo {author} {\bibfnamefont {P.~K.}\ \bibnamefont
  {Molony}}, \bibinfo {author} {\bibfnamefont {P.~D.}\ \bibnamefont {Gregory}},
  \bibinfo {author} {\bibfnamefont {Z.}~\bibnamefont {Ji}}, \bibinfo {author}
  {\bibfnamefont {B.}~\bibnamefont {Lu}}, \bibinfo {author} {\bibfnamefont
  {M.~P.}\ \bibnamefont {K\"oppinger}}, \bibinfo {author} {\bibfnamefont
  {C.~R.}\ \bibnamefont {{Le Sueur}}}, \bibinfo {author} {\bibfnamefont
  {C.~L.}\ \bibnamefont {Blackley}}, \bibinfo {author} {\bibfnamefont {J.~M.}\
  \bibnamefont {Hutson}}, \ and\ \bibinfo {author} {\bibfnamefont {S.~L.}\
  \bibnamefont {Cornish}},\ }\bibfield  {title} {\enquote {\bibinfo {title}
  {Creation of ultracold $^{87}${Rb}$^{133}${Cs} molecules in the rovibrational
  ground state},}\ }\href@noop {} {\bibfield  {journal} {\bibinfo  {journal}
  {Phys. Rev. Lett.}\ }\textbf {\bibinfo {volume} {113}},\ \bibinfo {pages}
  {255301} (\bibinfo {year} {2014})}\BibitemShut {NoStop}%
\bibitem [{\citenamefont {Frisch}\ \emph {et~al.}(2014)\citenamefont {Frisch},
  \citenamefont {Mark}, \citenamefont {Aikawa}, \citenamefont {Ferlaino},
  \citenamefont {Bohn}, \citenamefont {Makrides}, \citenamefont {Petrov},\ and\
  \citenamefont {Kotochigova}}]{Frisch:2014}%
  \BibitemOpen
  \bibfield  {author} {\bibinfo {author} {\bibfnamefont {A.}~\bibnamefont
  {Frisch}}, \bibinfo {author} {\bibfnamefont {M.}~\bibnamefont {Mark}},
  \bibinfo {author} {\bibfnamefont {K.}~\bibnamefont {Aikawa}}, \bibinfo
  {author} {\bibfnamefont {F.}~\bibnamefont {Ferlaino}}, \bibinfo {author}
  {\bibfnamefont {J.~L.}\ \bibnamefont {Bohn}}, \bibinfo {author}
  {\bibfnamefont {C.}~\bibnamefont {Makrides}}, \bibinfo {author}
  {\bibfnamefont {A.}~\bibnamefont {Petrov}}, \ and\ \bibinfo {author}
  {\bibfnamefont {S.}~\bibnamefont {Kotochigova}},\ }\bibfield  {title}
  {\enquote {\bibinfo {title} {Quantum chaos in ultracold collisions of
  gas-phase erbium atoms},}\ }\href@noop {} {\bibfield  {journal} {\bibinfo
  {journal} {Nature}\ }\textbf {\bibinfo {volume} {507}},\ \bibinfo {pages}
  {475} (\bibinfo {year} {2014})}\BibitemShut {NoStop}%
\bibitem [{\citenamefont {Mehta}(1991)}]{Mehta:rmt:1991}%
  \BibitemOpen
  \bibfield  {author} {\bibinfo {author} {\bibfnamefont {M.~L.}\ \bibnamefont
  {Mehta}},\ }\href@noop {} {\emph {\bibinfo {title} {Random Matrices}}},\
  \bibinfo {edition} {2nd}\ ed.\ (\bibinfo  {publisher} {Academic Press},\
  \bibinfo {year} {1991})\BibitemShut {NoStop}%
\bibitem [{\citenamefont {Guhr}\ \emph {et~al.}(1998)\citenamefont {Guhr},
  \citenamefont {M\"uller-Groeling},\ and\ \citenamefont
  {Weidenm\"uller}}]{Guhr:1998}%
  \BibitemOpen
  \bibfield  {author} {\bibinfo {author} {\bibfnamefont {T.}~\bibnamefont
  {Guhr}}, \bibinfo {author} {\bibfnamefont {A.}~\bibnamefont
  {M\"uller-Groeling}}, \ and\ \bibinfo {author} {\bibfnamefont {H.~A.}\
  \bibnamefont {Weidenm\"uller}},\ }\bibfield  {title} {\enquote {\bibinfo
  {title} {Random matrix theories in quantum physics: common concepts},}\
  }\href@noop {} {\bibfield  {journal} {\bibinfo  {journal} {Phys. Rep.}\
  }\textbf {\bibinfo {volume} {299}},\ \bibinfo {pages} {189} (\bibinfo {year}
  {1998})}\BibitemShut {NoStop}%
\bibitem [{\citenamefont {Wigner}(1955)}]{Wigner:1955}%
  \BibitemOpen
  \bibfield  {author} {\bibinfo {author} {\bibfnamefont {E.~P.}\ \bibnamefont
  {Wigner}},\ }\bibfield  {title} {\enquote {\bibinfo {title} {Characteristic
  vectors of bordered matrices with infinite dimensions},}\ }\href
  {http://www.jstor.org/stable/1970079} {\bibfield  {journal} {\bibinfo
  {journal} {Ann. Math.}\ }\bibinfo {series} {Second Series},\ \textbf
  {\bibinfo {volume} {62}},\ \bibinfo {pages} {pp. 548} (\bibinfo {year}
  {1955})}\BibitemShut {NoStop}%
\bibitem [{\citenamefont {Dyson}(1962)}]{Dyson:1962}%
  \BibitemOpen
  \bibfield  {author} {\bibinfo {author} {\bibfnamefont {F.~J.}\ \bibnamefont
  {Dyson}},\ }\bibfield  {title} {\enquote {\bibinfo {title} {Statistical
  theory of the energy levels of complex systems. {I}},}\ }\href {\doibase
  http://dx.doi.org/10.1063/1.1703773} {\bibfield  {journal} {\bibinfo
  {journal} {J. Math. Phys.}\ }\textbf {\bibinfo {volume} {3}},\ \bibinfo
  {pages} {140} (\bibinfo {year} {1962})}\BibitemShut {NoStop}%
\bibitem [{\citenamefont {Mayle}\ \emph {et~al.}(2012)\citenamefont {Mayle},
  \citenamefont {Ruzic},\ and\ \citenamefont {Bohn}}]{Mayle:2012}%
  \BibitemOpen
  \bibfield  {author} {\bibinfo {author} {\bibfnamefont {M.}~\bibnamefont
  {Mayle}}, \bibinfo {author} {\bibfnamefont {B.~P.}\ \bibnamefont {Ruzic}}, \
  and\ \bibinfo {author} {\bibfnamefont {J.~L.}\ \bibnamefont {Bohn}},\
  }\bibfield  {title} {\enquote {\bibinfo {title} {Statistical aspects of
  ultracold resonant scattering},}\ }\href {\doibase
  10.1103/PhysRevA.85.062712} {\bibfield  {journal} {\bibinfo  {journal} {Phys.
  Rev. A}\ }\textbf {\bibinfo {volume} {85}},\ \bibinfo {pages} {062712}
  (\bibinfo {year} {2012})}\BibitemShut {NoStop}%
\bibitem [{\citenamefont {Mayle}\ \emph {et~al.}(2013)\citenamefont {Mayle},
  \citenamefont {Qu\'em\'ener}, \citenamefont {Ruzic},\ and\ \citenamefont
  {Bohn}}]{Mayle:2013}%
  \BibitemOpen
  \bibfield  {author} {\bibinfo {author} {\bibfnamefont {M.}~\bibnamefont
  {Mayle}}, \bibinfo {author} {\bibfnamefont {G.}~\bibnamefont {Qu\'em\'ener}},
  \bibinfo {author} {\bibfnamefont {B.~P.}\ \bibnamefont {Ruzic}}, \ and\
  \bibinfo {author} {\bibfnamefont {J.~L.}\ \bibnamefont {Bohn}},\ }\bibfield
  {title} {\enquote {\bibinfo {title} {Scattering of ultracold molecules in the
  highly resonant regime},}\ }\href {\doibase 10.1103/PhysRevA.87.012709}
  {\bibfield  {journal} {\bibinfo  {journal} {Phys. Rev. A}\ }\textbf {\bibinfo
  {volume} {87}},\ \bibinfo {pages} {012709} (\bibinfo {year}
  {2013})}\BibitemShut {NoStop}%
\bibitem [{\citenamefont {Maier}\ \emph
  {et~al.}(2015{\natexlab{b}})\citenamefont {Maier}, \citenamefont {Kadau},
  \citenamefont {Schmitt}, \citenamefont {Wenzel}, \citenamefont
  {Ferrier-Barbut}, \citenamefont {Pfau}, \citenamefont {Frisch}, \citenamefont
  {Baier}, \citenamefont {Aikawa}, \citenamefont {Chomaz}, \citenamefont
  {Mark}, \citenamefont {Ferlaino}, \citenamefont {Makrides}, \citenamefont
  {Tiesinga}, \citenamefont {Petrov},\ and\ \citenamefont
  {Kotochigova}}]{Maier:ChaosErDy:2015}%
  \BibitemOpen
  \bibfield  {author} {\bibinfo {author} {\bibfnamefont {T.}~\bibnamefont
  {Maier}}, \bibinfo {author} {\bibfnamefont {H.}~\bibnamefont {Kadau}},
  \bibinfo {author} {\bibfnamefont {M.}~\bibnamefont {Schmitt}}, \bibinfo
  {author} {\bibfnamefont {M.}~\bibnamefont {Wenzel}}, \bibinfo {author}
  {\bibfnamefont {I.}~\bibnamefont {Ferrier-Barbut}}, \bibinfo {author}
  {\bibfnamefont {T.}~\bibnamefont {Pfau}}, \bibinfo {author} {\bibfnamefont
  {A.}~\bibnamefont {Frisch}}, \bibinfo {author} {\bibfnamefont
  {S.}~\bibnamefont {Baier}}, \bibinfo {author} {\bibfnamefont
  {K.}~\bibnamefont {Aikawa}}, \bibinfo {author} {\bibfnamefont
  {L.}~\bibnamefont {Chomaz}}, \bibinfo {author} {\bibfnamefont {M.~J.}\
  \bibnamefont {Mark}}, \bibinfo {author} {\bibfnamefont {F.}~\bibnamefont
  {Ferlaino}}, \bibinfo {author} {\bibfnamefont {C.}~\bibnamefont {Makrides}},
  \bibinfo {author} {\bibfnamefont {E.}~\bibnamefont {Tiesinga}}, \bibinfo
  {author} {\bibfnamefont {A.}~\bibnamefont {Petrov}}, \ and\ \bibinfo {author}
  {\bibfnamefont {S.}~\bibnamefont {Kotochigova}},\ }\bibfield  {title}
  {\enquote {\bibinfo {title} {Emergence of chaotic scattering in ultracold
  {Er} and {Dy}},}\ }\href {\doibase 10.1103/PhysRevX.5.041029} {\bibfield
  {journal} {\bibinfo  {journal} {Phys. Rev. X}\ }\textbf {\bibinfo {volume}
  {5}},\ \bibinfo {pages} {041029} (\bibinfo {year}
  {2015}{\natexlab{b}})}\BibitemShut {NoStop}%
\bibitem [{\citenamefont {Jachymski}\ and\ \citenamefont
  {Julienne}(2015)}]{Jachymski:ChaoticScat:2015}%
  \BibitemOpen
  \bibfield  {author} {\bibinfo {author} {\bibfnamefont {K.}~\bibnamefont
  {Jachymski}}\ and\ \bibinfo {author} {\bibfnamefont {P.~S.}\ \bibnamefont
  {Julienne}},\ }\bibfield  {title} {\enquote {\bibinfo {title} {Chaotic
  scattering in the presence of a dense set of overlapping {F}eshbach
  resonances},}\ }\href@noop {} {\bibfield  {journal} {\bibinfo  {journal}
  {Phys. Rev. A}\ }\textbf {\bibinfo {volume} {92}},\ \bibinfo {pages} {020702}
  (\bibinfo {year} {2015})}\BibitemShut {NoStop}%
\bibitem [{\citenamefont {Bohigas}\ \emph {et~al.}(1984)\citenamefont
  {Bohigas}, \citenamefont {Giannoni},\ and\ \citenamefont
  {Schmit}}]{Bohigas:1984}%
  \BibitemOpen
  \bibfield  {author} {\bibinfo {author} {\bibfnamefont {O.}~\bibnamefont
  {Bohigas}}, \bibinfo {author} {\bibfnamefont {M.~J.}\ \bibnamefont
  {Giannoni}}, \ and\ \bibinfo {author} {\bibfnamefont {C.}~\bibnamefont
  {Schmit}},\ }\bibfield  {title} {\enquote {\bibinfo {title} {Characterization
  of chaotic quantum spectra and universality of level fluctuation laws},}\
  }\href@noop {} {\bibfield  {journal} {\bibinfo  {journal} {Phys. Rev. Lett.}\
  }\textbf {\bibinfo {volume} {52}},\ \bibinfo {pages} {1} (\bibinfo {year}
  {1984})}\BibitemShut {NoStop}%
\bibitem [{\citenamefont {Mur-Petit}\ and\ \citenamefont
  {Molina}(2015)}]{Mur-Petit:2015}%
  \BibitemOpen
  \bibfield  {author} {\bibinfo {author} {\bibfnamefont {J.}~\bibnamefont
  {Mur-Petit}}\ and\ \bibinfo {author} {\bibfnamefont {R.~A.}\ \bibnamefont
  {Molina}},\ }\bibfield  {title} {\enquote {\bibinfo {title} {Spectral
  statistics of molecular resonances in erbium isotopes: How chaotic are
  they?}}\ }\href {\doibase 10.1103/PhysRevE.92.042906} {\bibfield  {journal}
  {\bibinfo  {journal} {Phys. Rev. E}\ }\textbf {\bibinfo {volume} {92}},\
  \bibinfo {pages} {042906} (\bibinfo {year} {2015})}\BibitemShut {NoStop}%
\bibitem [{\citenamefont {Green}\ \emph {et~al.}(2016)\citenamefont {Green},
  \citenamefont {Vaillant}, \citenamefont {Frye}, \citenamefont {Morita},\ and\
  \citenamefont {Hutson}}]{Green:chaos:2016}%
  \BibitemOpen
  \bibfield  {author} {\bibinfo {author} {\bibfnamefont {D.~G.}\ \bibnamefont
  {Green}}, \bibinfo {author} {\bibfnamefont {C.~L.}\ \bibnamefont {Vaillant}},
  \bibinfo {author} {\bibfnamefont {M.~D.}\ \bibnamefont {Frye}}, \bibinfo
  {author} {\bibfnamefont {M.}~\bibnamefont {Morita}}, \ and\ \bibinfo {author}
  {\bibfnamefont {J.~M.}\ \bibnamefont {Hutson}},\ }\bibfield  {title}
  {\enquote {\bibinfo {title} {Quantum chaos in ultracold collisions between
  $\text{Yb}({}^{1}\text{S}_{0})$ and $\text{Yb}({}^{3}\text{P}_{2})$},}\
  }\href {\doibase 10.1103/PhysRevA.93.022703} {\bibfield  {journal} {\bibinfo
  {journal} {Phys. Rev. A}\ }\textbf {\bibinfo {volume} {93}},\ \bibinfo
  {pages} {022703} (\bibinfo {year} {2016})}\BibitemShut {NoStop}%
\bibitem [{\citenamefont {Gonz\'alez-Mart\'{\i}nez}\ and\ \citenamefont
  {\.Zuchowski}(2015)}]{Gonzalez-Martinez:2015}%
  \BibitemOpen
  \bibfield  {author} {\bibinfo {author} {\bibfnamefont {M.~L.}\ \bibnamefont
  {Gonz\'alez-Mart\'{\i}nez}}\ and\ \bibinfo {author} {\bibfnamefont {P.~S.}\
  \bibnamefont {\.Zuchowski}},\ }\bibfield  {title} {\enquote {\bibinfo {title}
  {Magnetically tunable {F}eshbach resonances in {Li+Er}},}\ }\href {\doibase
  10.1103/PhysRevA.92.022708} {\bibfield  {journal} {\bibinfo  {journal} {Phys.
  Rev. A}\ }\textbf {\bibinfo {volume} {92}},\ \bibinfo {pages} {022708}
  (\bibinfo {year} {2015})}\BibitemShut {NoStop}%
\bibitem [{\citenamefont {Bernstein}(1979)}]{Bernstein:1979}%
  \BibitemOpen
  \bibinfo {editor} {\bibfnamefont {R.~B.}\ \bibnamefont {Bernstein}},\ ed.,\
  \href@noop {} {\emph {\bibinfo {title} {Atom-Molecule Collision Theory: a
  Guide for the Experimentalist}}}\ (\bibinfo  {publisher} {Plenum Press},\
  \bibinfo {address} {New York},\ \bibinfo {year} {1979})\BibitemShut {NoStop}%
\bibitem [{\citenamefont {Hutson}(1991)}]{Hutson:AMVCD:1991}%
  \BibitemOpen
  \bibfield  {author} {\bibinfo {author} {\bibfnamefont {J.~M.}\ \bibnamefont
  {Hutson}},\ }\bibfield  {title} {\enquote {\bibinfo {title} {An introduction
  to the dynamics of {V}an der {W}aals molecules},}\ }in\ \href@noop {} {\emph
  {\bibinfo {booktitle} {Advances in Molecular Vibrations and Collision
  Dynamics}}},\ Vol.~\bibinfo {volume} {1A}\ (\bibinfo  {publisher} {JAI
  Press},\ \bibinfo {address} {Greenwich, Connecticut},\ \bibinfo {year}
  {1991})\ pp.\ \bibinfo {pages} {1--45}\BibitemShut {NoStop}%
\bibitem [{\citenamefont {Hutson}(1994)}]{Hutson:CPC:1994}%
  \BibitemOpen
  \bibfield  {author} {\bibinfo {author} {\bibfnamefont {J.~M.}\ \bibnamefont
  {Hutson}},\ }\bibfield  {title} {\enquote {\bibinfo {title} {Coupled-channel
  methods for solving the bound-state {S}chr\"odinger equation},}\ }\href@noop
  {} {\bibfield  {journal} {\bibinfo  {journal} {Comput. Phys. Commun.}\
  }\textbf {\bibinfo {volume} {84}},\ \bibinfo {pages} {1} (\bibinfo {year}
  {1994})}\BibitemShut {NoStop}%
\bibitem [{\citenamefont {Hutson}\ and\ \citenamefont
  {Green}(1994)}]{molscat:v14}%
  \BibitemOpen
  \bibfield  {author} {\bibinfo {author} {\bibfnamefont {J.~M.}\ \bibnamefont
  {Hutson}}\ and\ \bibinfo {author} {\bibfnamefont {S.}~\bibnamefont {Green}},\
  }\href@noop {} {\enquote {\bibinfo {title} {{MOLSCAT} computer program,
  version 14},}\ }\bibinfo {howpublished} {distributed by Collaborative
  Computational Project No.\ 6 of the UK Engineering and Physical Sciences
  Research Council} (\bibinfo {year} {1994})\BibitemShut {NoStop}%
\bibitem [{\citenamefont {Hutson}(1993)}]{hutson:bound:1993}%
  \BibitemOpen
  \bibfield  {author} {\bibinfo {author} {\bibfnamefont {J.~M.}\ \bibnamefont
  {Hutson}},\ }\href@noop {} {\enquote {\bibinfo {title} {{BOUND} computer
  program, version 5},}\ }\bibinfo {howpublished} {distributed by Collaborative
  Computational Project No.\ 6 of the UK Engineering and Physical Sciences
  Research Council} (\bibinfo {year} {1993})\BibitemShut {NoStop}%
\bibitem [{\citenamefont {Hummon}\ \emph {et~al.}(2011)\citenamefont {Hummon},
  \citenamefont {Tscherbul}, \citenamefont {K\l{}os}, \citenamefont {Lu},
  \citenamefont {Tsikata}, \citenamefont {Campbell}, \citenamefont {Dalgarno},\
  and\ \citenamefont {Doyle}}]{Hummon:2011}%
  \BibitemOpen
  \bibfield  {author} {\bibinfo {author} {\bibfnamefont {M.~T.}\ \bibnamefont
  {Hummon}}, \bibinfo {author} {\bibfnamefont {T.~V.}\ \bibnamefont
  {Tscherbul}}, \bibinfo {author} {\bibfnamefont {J.}~\bibnamefont {K\l{}os}},
  \bibinfo {author} {\bibfnamefont {H.-I.}\ \bibnamefont {Lu}}, \bibinfo
  {author} {\bibfnamefont {E.}~\bibnamefont {Tsikata}}, \bibinfo {author}
  {\bibfnamefont {W.~C.}\ \bibnamefont {Campbell}}, \bibinfo {author}
  {\bibfnamefont {A.}~\bibnamefont {Dalgarno}}, \ and\ \bibinfo {author}
  {\bibfnamefont {J.~M.}\ \bibnamefont {Doyle}},\ }\bibfield  {title} {\enquote
  {\bibinfo {title} {Cold {N + NH} collisions in a magnetic trap},}\
  }\href@noop {} {\bibfield  {journal} {\bibinfo  {journal} {Phys. Rev. Lett.}\
  }\textbf {\bibinfo {volume} {106}},\ \bibinfo {pages} {053201} (\bibinfo
  {year} {2011})}\BibitemShut {NoStop}%
\bibitem [{\citenamefont {Lu}\ \emph {et~al.}(2014)\citenamefont {Lu},
  \citenamefont {Kozyryev}, \citenamefont {Hemmerling}, \citenamefont
  {Piskorski},\ and\ \citenamefont {Doyle}}]{Lu:CaF:2014}%
  \BibitemOpen
  \bibfield  {author} {\bibinfo {author} {\bibfnamefont {H.-I.}\ \bibnamefont
  {Lu}}, \bibinfo {author} {\bibfnamefont {I.}~\bibnamefont {Kozyryev}},
  \bibinfo {author} {\bibfnamefont {B.}~\bibnamefont {Hemmerling}}, \bibinfo
  {author} {\bibfnamefont {J.}~\bibnamefont {Piskorski}}, \ and\ \bibinfo
  {author} {\bibfnamefont {J.~M.}\ \bibnamefont {Doyle}},\ }\bibfield  {title}
  {\enquote {\bibinfo {title} {Magnetic trapping of molecules via optical
  loading and magnetic slowing},}\ }\href@noop {} {\bibfield  {journal}
  {\bibinfo  {journal} {Phys. Rev. Lett.}\ }\textbf {\bibinfo {volume} {112}},\
  \bibinfo {pages} {113006} (\bibinfo {year} {2014})}\BibitemShut {NoStop}%
\bibitem [{\citenamefont {Tscherbul}\ \emph {et~al.}(2011)\citenamefont
  {Tscherbul}, \citenamefont {K\l{}os},\ and\ \citenamefont
  {Buchachenko}}]{Tscherbul:2Sig:2011}%
  \BibitemOpen
  \bibfield  {author} {\bibinfo {author} {\bibfnamefont {T.~V.}\ \bibnamefont
  {Tscherbul}}, \bibinfo {author} {\bibfnamefont {J.}~\bibnamefont {K\l{}os}},
  \ and\ \bibinfo {author} {\bibfnamefont {A.~A.}\ \bibnamefont
  {Buchachenko}},\ }\bibfield  {title} {\enquote {\bibinfo {title} {Ultracold
  spin-polarized mixtures of {${}^{2}\Sigma$} molecules with {$S$}-state atoms:
  Collisional stability and implications for sympathetic cooling},}\ }\href
  {\doibase 10.1103/PhysRevA.84.040701} {\bibfield  {journal} {\bibinfo
  {journal} {Phys. Rev. A}\ }\textbf {\bibinfo {volume} {84}},\ \bibinfo
  {pages} {040701} (\bibinfo {year} {2011})}\BibitemShut {NoStop}%
\bibitem [{\citenamefont {Werner}\ \emph {et~al.}(2015)\citenamefont {Werner},
  \citenamefont {Knowles}, \citenamefont {Knizia}, \citenamefont {Manby},
  \citenamefont {{Sch\"{u}tz}}, \citenamefont {Celani}, \citenamefont {Korona},
  \citenamefont {Lindh}, \citenamefont {Mitrushenkov}, \citenamefont {Rauhut},
  \citenamefont {Shamasundar}, \citenamefont {Adler}, \citenamefont {Amos},
  \citenamefont {Bernhardsson}, \citenamefont {Berning}, \citenamefont
  {Cooper}, \citenamefont {Deegan}, \citenamefont {Dobbyn}, \citenamefont
  {Eckert}, \citenamefont {Goll}, \citenamefont {Hampel}, \citenamefont
  {Hesselmann}, \citenamefont {Hetzer}, \citenamefont {Hrenar}, \citenamefont
  {Jansen}, \citenamefont {K\"oppl}, \citenamefont {Liu}, \citenamefont
  {Lloyd}, \citenamefont {Mata}, \citenamefont {May}, \citenamefont
  {McNicholas}, \citenamefont {Meyer}, \citenamefont {Mura}, \citenamefont
  {Nicklass}, \citenamefont {O'Neill}, \citenamefont {Palmieri}, \citenamefont
  {Pfl\"uger}, \citenamefont {Pitzer}, \citenamefont {Reiher}, \citenamefont
  {Shiozaki}, \citenamefont {Stoll}, \citenamefont {Stone}, \citenamefont
  {Tarroni}, \citenamefont {Thorsteinsson}, \citenamefont {Wang},\ and\
  \citenamefont {Wolf}}]{MOLPRO:2010.1}%
  \BibitemOpen
  \bibfield  {author} {\bibinfo {author} {\bibfnamefont {H.-J.}\ \bibnamefont
  {Werner}}, \bibinfo {author} {\bibfnamefont {P.~J.}\ \bibnamefont {Knowles}},
  \bibinfo {author} {\bibfnamefont {G.}~\bibnamefont {Knizia}}, \bibinfo
  {author} {\bibfnamefont {F.~R.}\ \bibnamefont {Manby}}, \bibinfo {author}
  {\bibfnamefont {M.}~\bibnamefont {{Sch\"{u}tz}}}, \bibinfo {author}
  {\bibfnamefont {P.}~\bibnamefont {Celani}}, \bibinfo {author} {\bibfnamefont
  {T.}~\bibnamefont {Korona}}, \bibinfo {author} {\bibfnamefont
  {R.}~\bibnamefont {Lindh}}, \bibinfo {author} {\bibfnamefont
  {A.}~\bibnamefont {Mitrushenkov}}, \bibinfo {author} {\bibfnamefont
  {G.}~\bibnamefont {Rauhut}}, \bibinfo {author} {\bibfnamefont {K.~R.}\
  \bibnamefont {Shamasundar}}, \bibinfo {author} {\bibfnamefont {T.~B.}\
  \bibnamefont {Adler}}, \bibinfo {author} {\bibfnamefont {R.~D.}\ \bibnamefont
  {Amos}}, \bibinfo {author} {\bibfnamefont {A.}~\bibnamefont {Bernhardsson}},
  \bibinfo {author} {\bibfnamefont {A.}~\bibnamefont {Berning}}, \bibinfo
  {author} {\bibfnamefont {D.~L.}\ \bibnamefont {Cooper}}, \bibinfo {author}
  {\bibfnamefont {M.~J.~O.}\ \bibnamefont {Deegan}}, \bibinfo {author}
  {\bibfnamefont {A.~J.}\ \bibnamefont {Dobbyn}}, \bibinfo {author}
  {\bibfnamefont {F.}~\bibnamefont {Eckert}}, \bibinfo {author} {\bibfnamefont
  {E.}~\bibnamefont {Goll}}, \bibinfo {author} {\bibfnamefont {C.}~\bibnamefont
  {Hampel}}, \bibinfo {author} {\bibfnamefont {A.}~\bibnamefont {Hesselmann}},
  \bibinfo {author} {\bibfnamefont {G.}~\bibnamefont {Hetzer}}, \bibinfo
  {author} {\bibfnamefont {T.}~\bibnamefont {Hrenar}}, \bibinfo {author}
  {\bibfnamefont {G.}~\bibnamefont {Jansen}}, \bibinfo {author} {\bibfnamefont
  {C.}~\bibnamefont {K\"oppl}}, \bibinfo {author} {\bibfnamefont
  {Y.}~\bibnamefont {Liu}}, \bibinfo {author} {\bibfnamefont {A.~W.}\
  \bibnamefont {Lloyd}}, \bibinfo {author} {\bibfnamefont {R.~A.}\ \bibnamefont
  {Mata}}, \bibinfo {author} {\bibfnamefont {A.~J.}\ \bibnamefont {May}},
  \bibinfo {author} {\bibfnamefont {S.~J.}\ \bibnamefont {McNicholas}},
  \bibinfo {author} {\bibfnamefont {W.}~\bibnamefont {Meyer}}, \bibinfo
  {author} {\bibfnamefont {M.~E.}\ \bibnamefont {Mura}}, \bibinfo {author}
  {\bibfnamefont {A.}~\bibnamefont {Nicklass}}, \bibinfo {author}
  {\bibfnamefont {D.~P.}\ \bibnamefont {O'Neill}}, \bibinfo {author}
  {\bibfnamefont {P.}~\bibnamefont {Palmieri}}, \bibinfo {author}
  {\bibfnamefont {K.}~\bibnamefont {Pfl\"uger}}, \bibinfo {author}
  {\bibfnamefont {R.}~\bibnamefont {Pitzer}}, \bibinfo {author} {\bibfnamefont
  {M.}~\bibnamefont {Reiher}}, \bibinfo {author} {\bibfnamefont
  {T.}~\bibnamefont {Shiozaki}}, \bibinfo {author} {\bibfnamefont
  {H.}~\bibnamefont {Stoll}}, \bibinfo {author} {\bibfnamefont {A.~J.}\
  \bibnamefont {Stone}}, \bibinfo {author} {\bibfnamefont {R.}~\bibnamefont
  {Tarroni}}, \bibinfo {author} {\bibfnamefont {T.}~\bibnamefont
  {Thorsteinsson}}, \bibinfo {author} {\bibfnamefont {M.}~\bibnamefont {Wang}},
  \ and\ \bibinfo {author} {\bibfnamefont {A.}~\bibnamefont {Wolf}},\
  }\href@noop {} {\enquote {\bibinfo {title} {{MOLPRO}, version 2010.1, a
  package of {\em ab initio} programs},}\ } (\bibinfo {year} {2015}),\ \bibinfo
  {note} {see http://www.molpro.net}\BibitemShut {NoStop}%
\bibitem [{\citenamefont {Dunning}(1989)}]{Dunning:1989}%
  \BibitemOpen
  \bibfield  {author} {\bibinfo {author} {\bibfnamefont {T.~H.}\ \bibnamefont
  {Dunning}, \bibfnamefont {Jr.}},\ }\bibfield  {title} {\enquote {\bibinfo
  {title} {Gaussian basis sets for use in correlated molecular calculations.
  {I}. {T}he atoms boron through neon and hydrogen},}\ }\href@noop {}
  {\bibfield  {journal} {\bibinfo  {journal} {J. Chem. Phys.}\ }\textbf
  {\bibinfo {volume} {90}},\ \bibinfo {pages} {1007} (\bibinfo {year}
  {1989})}\BibitemShut {NoStop}%
\bibitem [{\citenamefont {Kendall}\ \emph {et~al.}(1992)\citenamefont
  {Kendall}, \citenamefont {Dunning},\ and\ \citenamefont
  {Harrison}}]{Kendall:1992}%
  \BibitemOpen
  \bibfield  {author} {\bibinfo {author} {\bibfnamefont {R.~A.}\ \bibnamefont
  {Kendall}}, \bibinfo {author} {\bibfnamefont {T.~H.}\ \bibnamefont
  {Dunning}}, \ and\ \bibinfo {author} {\bibfnamefont {R.~J.}\ \bibnamefont
  {Harrison}},\ }\bibfield  {title} {\enquote {\bibinfo {title} {Electron
  affinities of the first-row atoms revisited. {S}ystematic basis sets and wave
  functions},}\ }\href@noop {} {\bibfield  {journal} {\bibinfo  {journal} {J.
  Chem. Phys.}\ }\textbf {\bibinfo {volume} {96}},\ \bibinfo {pages} {6796}
  (\bibinfo {year} {1992})}\BibitemShut {NoStop}%
\bibitem [{\citenamefont {Koput}\ and\ \citenamefont
  {Peterson}(2002)}]{Koput:2002}%
  \BibitemOpen
  \bibfield  {author} {\bibinfo {author} {\bibfnamefont {J.}~\bibnamefont
  {Koput}}\ and\ \bibinfo {author} {\bibfnamefont {K.~A.}\ \bibnamefont
  {Peterson}},\ }\bibfield  {title} {\enquote {\bibinfo {title} {Ab initio
  potential energy surface and vibrational-rotational energy levels of
  {X$^2\Sigma^+$} {CaOH}},}\ }\href@noop {} {\bibfield  {journal} {\bibinfo
  {journal} {J. Phys. Chem. A}\ }\textbf {\bibinfo {volume} {106}},\ \bibinfo
  {pages} {9595} (\bibinfo {year} {2002})}\BibitemShut {NoStop}%
\bibitem [{\citenamefont {Boys}\ and\ \citenamefont
  {Bernardi}(1970)}]{Boys:1970}%
  \BibitemOpen
  \bibfield  {author} {\bibinfo {author} {\bibfnamefont {S.~F.}\ \bibnamefont
  {Boys}}\ and\ \bibinfo {author} {\bibfnamefont {F.}~\bibnamefont
  {Bernardi}},\ }\bibfield  {title} {\enquote {\bibinfo {title} {The
  calculation of small molecular interactions by the differences of separate
  total energies. {S}ome procedures with reduced errors},}\ }\href@noop {}
  {\bibfield  {journal} {\bibinfo  {journal} {Mol. Phys.}\ }\textbf {\bibinfo
  {volume} {19}},\ \bibinfo {pages} {553} (\bibinfo {year} {1970})}\BibitemShut
  {NoStop}%
\bibitem [{\citenamefont {Kaledin}\ \emph {et~al.}(1999)\citenamefont
  {Kaledin}, \citenamefont {Bloch}, \citenamefont {McCarthy},\ and\
  \citenamefont {Field}}]{Kaledin:1999}%
  \BibitemOpen
  \bibfield  {author} {\bibinfo {author} {\bibfnamefont {L.~A.}\ \bibnamefont
  {Kaledin}}, \bibinfo {author} {\bibfnamefont {J.~C.}\ \bibnamefont {Bloch}},
  \bibinfo {author} {\bibfnamefont {M.~C.}\ \bibnamefont {McCarthy}}, \ and\
  \bibinfo {author} {\bibfnamefont {R.~W.}\ \bibnamefont {Field}},\ }\bibfield
  {title} {\enquote {\bibinfo {title} {Analysis and deperturbation of the
  {$A^2\Pi$} and {$B^2\Sigma^+$} states of {CaF}},}\ }\href@noop {} {\bibfield
  {journal} {\bibinfo  {journal} {J. Mol. Spectrosc.}\ }\textbf {\bibinfo
  {volume} {197}},\ \bibinfo {pages} {289 } (\bibinfo {year}
  {1999})}\BibitemShut {NoStop}%
\bibitem [{\citenamefont {Brody}\ \emph {et~al.}(1981)\citenamefont {Brody},
  \citenamefont {Flores}, \citenamefont {French}, \citenamefont {Mellow},
  \citenamefont {Pandey},\ and\ \citenamefont {Wong}}]{Brody:rmtreview:1981}%
  \BibitemOpen
  \bibfield  {author} {\bibinfo {author} {\bibfnamefont {T.~A.}\ \bibnamefont
  {Brody}}, \bibinfo {author} {\bibfnamefont {J.}~\bibnamefont {Flores}},
  \bibinfo {author} {\bibfnamefont {J.~B.}\ \bibnamefont {French}}, \bibinfo
  {author} {\bibfnamefont {P.~A.}\ \bibnamefont {Mellow}}, \bibinfo {author}
  {\bibfnamefont {A.}~\bibnamefont {Pandey}}, \ and\ \bibinfo {author}
  {\bibfnamefont {S.~S.~M.}\ \bibnamefont {Wong}},\ }\bibfield  {title}
  {\enquote {\bibinfo {title} {Random-matrix physics: spectrum and strength
  fluctuations},}\ }\href@noop {} {\bibfield  {journal} {\bibinfo  {journal}
  {Rev. Mod. Phys.}\ }\textbf {\bibinfo {volume} {53}},\ \bibinfo {pages} {385}
  (\bibinfo {year} {1981})}\BibitemShut {NoStop}%
\bibitem [{\citenamefont {Cvita\v{s}}\ \emph {et~al.}(2007)\citenamefont
  {Cvita\v{s}}, \citenamefont {Sold\'{a}n}, \citenamefont {Hutson},
  \citenamefont {Honvault},\ and\ \citenamefont {Launay}}]{Cvitas:li3:2007}%
  \BibitemOpen
  \bibfield  {author} {\bibinfo {author} {\bibfnamefont {M.~T.}\ \bibnamefont
  {Cvita\v{s}}}, \bibinfo {author} {\bibfnamefont {P.}~\bibnamefont
  {Sold\'{a}n}}, \bibinfo {author} {\bibfnamefont {J.~M.}\ \bibnamefont
  {Hutson}}, \bibinfo {author} {\bibfnamefont {P.}~\bibnamefont {Honvault}}, \
  and\ \bibinfo {author} {\bibfnamefont {J.~M.}\ \bibnamefont {Launay}},\
  }\bibfield  {title} {\enquote {\bibinfo {title} {Interactions and dynamics in
  {Li + Li}$_2$ ultracold collisions},}\ }\href@noop {} {\bibfield  {journal}
  {\bibinfo  {journal} {J. Chem. Phys.}\ }\textbf {\bibinfo {volume} {127}},\
  \bibinfo {pages} {074302} (\bibinfo {year} {2007})}\BibitemShut {NoStop}%
\bibitem [{\citenamefont {Wallis}\ and\ \citenamefont
  {Hutson}(2009)}]{Wallis:MgNH:2009}%
  \BibitemOpen
  \bibfield  {author} {\bibinfo {author} {\bibfnamefont {A.~O.~G.}\
  \bibnamefont {Wallis}}\ and\ \bibinfo {author} {\bibfnamefont {J.~M.}\
  \bibnamefont {Hutson}},\ }\bibfield  {title} {\enquote {\bibinfo {title}
  {Production of ultracold {NH} molecules by sympathetic cooling with {Mg}},}\
  }\href@noop {} {\bibfield  {journal} {\bibinfo  {journal} {Phys. Rev. Lett.}\
  }\textbf {\bibinfo {volume} {103}},\ \bibinfo {pages} {183201} (\bibinfo
  {year} {2009})}\BibitemShut {NoStop}%
\bibitem [{\citenamefont {Wallis}\ \emph {et~al.}(2011)\citenamefont {Wallis},
  \citenamefont {Longdon}, \citenamefont {\.Zuchowski},\ and\ \citenamefont
  {Hutson}}]{Wallis:LiNH:2011}%
  \BibitemOpen
  \bibfield  {author} {\bibinfo {author} {\bibfnamefont {A.~O.~G.}\
  \bibnamefont {Wallis}}, \bibinfo {author} {\bibfnamefont {E.~J.~J.}\
  \bibnamefont {Longdon}}, \bibinfo {author} {\bibfnamefont {P.~S.}\
  \bibnamefont {\.Zuchowski}}, \ and\ \bibinfo {author} {\bibfnamefont {J.~M.}\
  \bibnamefont {Hutson}},\ }\bibfield  {title} {\enquote {\bibinfo {title} {The
  prospects of sympathetic cooling of {NH} molecules with {Li} atoms},}\
  }\href@noop {} {\bibfield  {journal} {\bibinfo  {journal} {Eur. Phys. J. D}\
  }\textbf {\bibinfo {volume} {65}},\ \bibinfo {pages} {151} (\bibinfo {year}
  {2011})}\BibitemShut {NoStop}%
\bibitem [{Note1()}]{Note1}%
  \BibitemOpen
  \bibinfo {note} {Note that even $\lambda =1$ should not be interpreted as the
  true potential because there are significant uncertainties in the calculated
  potentials.}\BibitemShut {Stop}%
\bibitem [{\citenamefont {Hutson}(2011)}]{Hutson:field:2011}%
  \BibitemOpen
  \bibfield  {author} {\bibinfo {author} {\bibfnamefont {J.~M.}\ \bibnamefont
  {Hutson}},\ }\href@noop {} {\enquote {\bibinfo {title} {{FIELD} computer
  program, version 1},}\ } (\bibinfo {year} {2011})\BibitemShut {NoStop}%
\bibitem [{\citenamefont {Alexander}\ and\ \citenamefont
  {Manolopoulos}(1987)}]{Alexander:1987}%
  \BibitemOpen
  \bibfield  {author} {\bibinfo {author} {\bibfnamefont {M.~H.}\ \bibnamefont
  {Alexander}}\ and\ \bibinfo {author} {\bibfnamefont {D.~E.}\ \bibnamefont
  {Manolopoulos}},\ }\bibfield  {title} {\enquote {\bibinfo {title} {A stable
  linear reference potential algorithm for solution of the quantum
  close-coupled equations in molecular scattering theory},}\ }\href@noop {}
  {\bibfield  {journal} {\bibinfo  {journal} {J. Chem. Phys.}\ }\textbf
  {\bibinfo {volume} {86}},\ \bibinfo {pages} {2044} (\bibinfo {year}
  {1987})}\BibitemShut {NoStop}%
\bibitem [{\citenamefont {Mizushima}(1975)}]{Mizushima}%
  \BibitemOpen
  \bibfield  {author} {\bibinfo {author} {\bibfnamefont {M.}~\bibnamefont
  {Mizushima}},\ }\href@noop {} {\emph {\bibinfo {title} {Theory of Rotating
  Diatomic Molecules}}}\ (\bibinfo  {publisher} {Wiley},\ \bibinfo {address}
  {New York},\ \bibinfo {year} {1975})\BibitemShut {NoStop}%
\bibitem [{Note2()}]{Note2}%
  \BibitemOpen
  \bibinfo {note} {The anisotropy considered in Ref.\ \cite
  {Maier:ChaosErDy:2015} is based entirely on dispersion effects \cite
  {Petrov:2012}. The spread of dispersion coefficients for different potential
  curves of Er-Er is 10\% of their mean value, and that for Dy-Dy is 9\% of
  their mean value.}\BibitemShut {Stop}%
\bibitem [{Note3()}]{Note3}%
  \BibitemOpen
  \bibinfo {note} {We observe only a fraction of the total number of levels in
  the system, so $n \ll N$.}\BibitemShut {Stop}%
\bibitem [{\citenamefont {Berry}\ and\ \citenamefont
  {Robnik}(1984)}]{Berry:semiclassical:1984}%
  \BibitemOpen
  \bibfield  {author} {\bibinfo {author} {\bibfnamefont {M.~V.}\ \bibnamefont
  {Berry}}\ and\ \bibinfo {author} {\bibfnamefont {M.}~\bibnamefont {Robnik}},\
  }\bibfield  {title} {\enquote {\bibinfo {title} {Semiclassical level spacings
  when regular and chaotic orbits coexist},}\ }\href@noop {} {\bibfield
  {journal} {\bibinfo  {journal} {J. Phys. A}\ }\textbf {\bibinfo {volume}
  {17}},\ \bibinfo {pages} {2413} (\bibinfo {year} {1984})}\BibitemShut
  {NoStop}%
\bibitem [{\citenamefont {Izrailev}(1988)}]{Izrailev:1988}%
  \BibitemOpen
  \bibfield  {author} {\bibinfo {author} {\bibfnamefont {F.}~\bibnamefont
  {Izrailev}},\ }\bibfield  {title} {\enquote {\bibinfo {title} {Quantum
  localization and statistics of quasienergy spectrum in a classically chaotic
  system},}\ }\href {\doibase http://dx.doi.org/10.1016/0375-9601(88)90538-5}
  {\bibfield  {journal} {\bibinfo  {journal} {Phys. Lett. A}\ }\textbf
  {\bibinfo {volume} {134}},\ \bibinfo {pages} {13 } (\bibinfo {year}
  {1988})}\BibitemShut {NoStop}%
\bibitem [{\citenamefont {Brody}(1973)}]{Brody:1973}%
  \BibitemOpen
  \bibfield  {author} {\bibinfo {author} {\bibfnamefont {T.}~\bibnamefont
  {Brody}},\ }\bibfield  {title} {\enquote {\bibinfo {title} {A statistical
  measure for the repulsion of energy levels},}\ }\href@noop {} {\bibfield
  {journal} {\bibinfo  {journal} {Lett. Nuovo Cimento}\ }\textbf {\bibinfo
  {volume} {7}},\ \bibinfo {pages} {482} (\bibinfo {year} {1973})}\BibitemShut
  {NoStop}%
\bibitem [{\citenamefont {Barlow}(1989)}]{Barlow:1989}%
  \BibitemOpen
  \bibfield  {author} {\bibinfo {author} {\bibfnamefont {R.~J.}\ \bibnamefont
  {Barlow}},\ }\href@noop {} {\emph {\bibinfo {title} {Statistics: A Guide to
  the Use of Statistical Methods in the Physical Sciences}}}\ (\bibinfo
  {publisher} {Wiley},\ \bibinfo {year} {1989})\BibitemShut {NoStop}%
\bibitem [{\citenamefont {Berry}\ and\ \citenamefont
  {Tabor}(1977)}]{Berry:1977}%
  \BibitemOpen
  \bibfield  {author} {\bibinfo {author} {\bibfnamefont {M.~V.}\ \bibnamefont
  {Berry}}\ and\ \bibinfo {author} {\bibfnamefont {M.}~\bibnamefont {Tabor}},\
  }\bibfield  {title} {\enquote {\bibinfo {title} {Level clustering in the
  regular spectrum},}\ }\href {\doibase 10.1098/rspa.1977.0140} {\bibfield
  {journal} {\bibinfo  {journal} {Proc. R. Soc. A}\ }\textbf {\bibinfo {volume}
  {356}},\ \bibinfo {pages} {375} (\bibinfo {year} {1977})}\BibitemShut
  {NoStop}%
\bibitem [{\citenamefont {Rosenzweig}\ and\ \citenamefont
  {Porter}(1960)}]{Rosenzweig:levelrepulsion:1960}%
  \BibitemOpen
  \bibfield  {author} {\bibinfo {author} {\bibfnamefont {N.}~\bibnamefont
  {Rosenzweig}}\ and\ \bibinfo {author} {\bibfnamefont {C.~E.}\ \bibnamefont
  {Porter}},\ }\bibfield  {title} {\enquote {\bibinfo {title} {`{R}epulsion of
  energy levels' in complex atomic spectra},}\ }\href@noop {} {\bibfield
  {journal} {\bibinfo  {journal} {Phys. Rev.}\ }\textbf {\bibinfo {volume}
  {120}},\ \bibinfo {pages} {1698} (\bibinfo {year} {1960})}\BibitemShut
  {NoStop}%
\bibitem [{Note4()}]{Note4}%
  \BibitemOpen
  \bibinfo {note} {Although visually there may seem to be relatively few steep
  states for Li+CaF, counting them with respect to $\lambda $ at fixed energy
  reveals that approximately 1/3 of the levels are of this type.}\BibitemShut
  {Stop}%
\bibitem [{\citenamefont {Holmgren}\ \emph {et~al.}(1977)\citenamefont
  {Holmgren}, \citenamefont {Waldman},\ and\ \citenamefont
  {Klemperer}}]{Holm77}%
  \BibitemOpen
  \bibfield  {author} {\bibinfo {author} {\bibfnamefont {S.~L.}\ \bibnamefont
  {Holmgren}}, \bibinfo {author} {\bibfnamefont {M.}~\bibnamefont {Waldman}}, \
  and\ \bibinfo {author} {\bibfnamefont {W.}~\bibnamefont {Klemperer}},\
  }\bibfield  {title} {\enquote {\bibinfo {title} {Internal dynamics of van der
  {W}aals complexes. {I}. {B}orn--{O}ppenheimer separation of radial and
  angular motion},}\ }\href@noop {} {\bibfield  {journal} {\bibinfo  {journal}
  {J. Chem. Phys.}\ }\textbf {\bibinfo {volume} {67}},\ \bibinfo {pages} {4414}
  (\bibinfo {year} {1977})}\BibitemShut {NoStop}%
\bibitem [{\citenamefont {Hutson}\ and\ \citenamefont
  {Howard}(1980)}]{Hutson:CBO:1980}%
  \BibitemOpen
  \bibfield  {author} {\bibinfo {author} {\bibfnamefont {J.~M.}\ \bibnamefont
  {Hutson}}\ and\ \bibinfo {author} {\bibfnamefont {B.~J.}\ \bibnamefont
  {Howard}},\ }\bibfield  {title} {\enquote {\bibinfo {title} {Spectroscopic
  properties and potential surfaces for atom-diatom van der {W}aals
  molecules},}\ }\href@noop {} {\bibfield  {journal} {\bibinfo  {journal} {Mol.
  Phys.}\ }\textbf {\bibinfo {volume} {41}},\ \bibinfo {pages} {1123} (\bibinfo
  {year} {1980})}\BibitemShut {NoStop}%
\bibitem [{\citenamefont {Hutson}\ and\ \citenamefont
  {Howard}(1982)}]{H82RgHCl}%
  \BibitemOpen
  \bibfield  {author} {\bibinfo {author} {\bibfnamefont {J.~M.}\ \bibnamefont
  {Hutson}}\ and\ \bibinfo {author} {\bibfnamefont {B.~J.}\ \bibnamefont
  {Howard}},\ }\bibfield  {title} {\enquote {\bibinfo {title} {Anisotropic
  intermolecular forces. {I}. {R}are gas-hydrogen chloride systems},}\
  }\href@noop {} {\bibfield  {journal} {\bibinfo  {journal} {Mol. Phys.}\
  }\textbf {\bibinfo {volume} {45}},\ \bibinfo {pages} {769} (\bibinfo {year}
  {1982})}\BibitemShut {NoStop}%
\bibitem [{\citenamefont {{Le Sueur}}\ \emph {et~al.}(1993)\citenamefont {{Le
  Sueur}}, \citenamefont {Henderson},\ and\ \citenamefont
  {Tennyson}}]{LeSueur:1993}%
  \BibitemOpen
  \bibfield  {author} {\bibinfo {author} {\bibfnamefont {C.~R.}\ \bibnamefont
  {{Le Sueur}}}, \bibinfo {author} {\bibfnamefont {J.~R.}\ \bibnamefont
  {Henderson}}, \ and\ \bibinfo {author} {\bibfnamefont {J.}~\bibnamefont
  {Tennyson}},\ }\bibfield  {title} {\enquote {\bibinfo {title} {Gateway states
  and bath states in the vibrational spectrum of {H}$^+_3$},}\ }\href {\doibase
  http://dx.doi.org/10.1016/0009-2614(93)80163-J} {\bibfield  {journal}
  {\bibinfo  {journal} {Chem. Phys. Lett.}\ }\textbf {\bibinfo {volume}
  {206}},\ \bibinfo {pages} {429 } (\bibinfo {year} {1993})}\BibitemShut
  {NoStop}%
\bibitem [{\citenamefont {de~Polavieja}\ \emph {et~al.}(1994)\citenamefont
  {de~Polavieja}, \citenamefont {Borondo},\ and\ \citenamefont
  {Benito}}]{DePolavieja:1994}%
  \BibitemOpen
  \bibfield  {author} {\bibinfo {author} {\bibfnamefont {G.~G.}\ \bibnamefont
  {de~Polavieja}}, \bibinfo {author} {\bibfnamefont {F.}~\bibnamefont
  {Borondo}}, \ and\ \bibinfo {author} {\bibfnamefont {R.~M.}\ \bibnamefont
  {Benito}},\ }\bibfield  {title} {\enquote {\bibinfo {title} {Scars in groups
  of eigenstates in a classically chaotic system},}\ }\href@noop {} {\bibfield
  {journal} {\bibinfo  {journal} {Phys. Rev. Lett.}\ }\textbf {\bibinfo
  {volume} {73}},\ \bibinfo {pages} {1613} (\bibinfo {year}
  {1994})}\BibitemShut {NoStop}%
\bibitem [{\citenamefont {Wright}\ and\ \citenamefont
  {Hutson}(1999)}]{Wright:1999}%
  \BibitemOpen
  \bibfield  {author} {\bibinfo {author} {\bibfnamefont {N.~J.}\ \bibnamefont
  {Wright}}\ and\ \bibinfo {author} {\bibfnamefont {J.~M.}\ \bibnamefont
  {Hutson}},\ }\bibfield  {title} {\enquote {\bibinfo {title} {Regular and
  irregular vibrational states: Localized anharmonic modes in {Ar}$_3$},}\
  }\href {\doibase http://dx.doi.org/10.1063/1.478057} {\bibfield  {journal}
  {\bibinfo  {journal} {J. Chem. Phys.}\ }\textbf {\bibinfo {volume} {110}},\
  \bibinfo {pages} {902} (\bibinfo {year} {1999})}\BibitemShut {NoStop}%
\bibitem [{\citenamefont {Wright}\ and\ \citenamefont
  {Hutson}(2000)}]{Wright:2000}%
  \BibitemOpen
  \bibfield  {author} {\bibinfo {author} {\bibfnamefont {N.~J.}\ \bibnamefont
  {Wright}}\ and\ \bibinfo {author} {\bibfnamefont {J.~M.}\ \bibnamefont
  {Hutson}},\ }\bibfield  {title} {\enquote {\bibinfo {title} {Regular and
  irregular vibrational states: Localized anharmonic modes and transition-state
  spectroscopy of {Na}$_3$},}\ }\href {\doibase
  http://dx.doi.org/10.1063/1.480905} {\bibfield  {journal} {\bibinfo
  {journal} {J. Chem. Phys.}\ }\textbf {\bibinfo {volume} {112}},\ \bibinfo
  {pages} {3214} (\bibinfo {year} {2000})}\BibitemShut {NoStop}%
\bibitem [{\citenamefont {Petrov}\ \emph {et~al.}(2012)\citenamefont {Petrov},
  \citenamefont {Tiesinga},\ and\ \citenamefont {Kotochigova}}]{Petrov:2012}%
  \BibitemOpen
  \bibfield  {author} {\bibinfo {author} {\bibfnamefont {A.}~\bibnamefont
  {Petrov}}, \bibinfo {author} {\bibfnamefont {E.}~\bibnamefont {Tiesinga}}, \
  and\ \bibinfo {author} {\bibfnamefont {S.}~\bibnamefont {Kotochigova}},\
  }\bibfield  {title} {\enquote {\bibinfo {title} {Anisotropy-induced
  {F}eshbach resonances in a quantum dipolar gas of highly magnetic atoms},}\
  }\href {\doibase 10.1103/PhysRevLett.109.103002} {\bibfield  {journal}
  {\bibinfo  {journal} {Phys. Rev. Lett.}\ }\textbf {\bibinfo {volume} {109}},\
  \bibinfo {pages} {103002} (\bibinfo {year} {2012})}\BibitemShut {NoStop}%
\end{thebibliography}%
\end{document}